\documentclass[structabstract]{aa}  

\usepackage{graphicx}
\usepackage{txfonts}
\usepackage{multirow}
\usepackage[authoryear]{natbib}

\begin{document}
   \title{X-ray view of four high--luminosity Swift/BAT AGN: Unveiling obscuration and reflection with Suzaku}

   \author{V. Fioretti
          \inst{1}\inst{,2}\fnmsep\thanks{e-mail: fioretti@iasfbo.inaf.it},
          L. Angelini
          \inst{3},
          R.F. Mushotzky
          \inst{3},
          M. Koss
          \inst{3},
          \and
          G. Malaguti
          \inst{1}
          }

   \institute{INAF/Istituto di Astrofisica Spaziale e Fisica cosmica, via Gobetti 101, 40129 Bologna, Italy\\
              \and
              Dipartimento di Astronomia, Universit\`a di Bologna, via Ranzani 1, 40127 Bologna, Italy \\
              \and
              NASA/Goddard Space Flight Center, Greenbelt, MD 20771, USA
             }

   \date{Received XXXXXXX, XXXX; accepted XXXXXXX, XXXX}

 
  \abstract
   {}
{A complete census of obscured Active Galactic Nuclei (AGN) is necessary to reveal the history of the super massive black hole (SMBH) growth and galaxy evolution in the Universe given the complex feedback processes and the fact that much of this growth occurs in an obscured phase. In this context, hard X-ray surveys and dedicated follow-up observations represent a unique tool for selecting highly absorbed AGN and for characterizing the obscuring matter surrounding the SMBH. 
Here we focus on the absorption and reflection occurring in highly luminous, quasar-like AGN, to study the relation between the geometry of the absorbing matter and the AGN nature (e.g. X-ray, optical, and radio properties), and to help to determine the column density dependency on the AGN luminosity.}
{The Swift/BAT nine-month survey observed 153 AGN, all with ultra-hard X-ray BAT fluxes in excess of $10^{-11}$ erg cm$^{-2}$ s$^{-1}$ and an average redshift of 0.03. Among them, four of the most luminous BAT AGN ($44.73\rm<LogL_{\rm BAT}<45.31$) were selected as targets of Suzaku follow-up observations: J2246.0+3941 (3C 452), J0407.4+0339 (3C 105), J0318.7+6828, and J0918.5+0425. 
The column density, scattered/reflected emission, the properties of the Fe K line, and a possible variability are fully analyzed. For the latter, the spectral properties from Chandra, XMM-Newton and Swift/XRT public observations were compared with the present Suzaku analysis, adding an original spectral analysis when non was available from the literature. 
}
{Of our sample, 3C 452 is the only certain Compton-thick AGN candidate because of i) the high absorption (N$_{\rm H}\sim4\times10^{23}$ cm$^{-2}$) and strong Compton reflection; ii) the lack of variability; iii) the ''buried'' nature, i.e. the low scattering fraction ($<0.5\%$) and the extremely low relative [OIII] luminosity. In contrast 3C 105 is not reflection-dominated, despite the comparable column density, X-ray luminosity and radio morphology, but shows a strong long-term variability in flux and scattering fraction, consistent with the soft emission being scattered from a distant region (e.g., the narrow emission line region).
The sample presents high ($>100$) X-to-[OIII] luminosity ratios, with an extreme value of R$^{\rm X}_{\rm [OIII]}\sim800$ for 3C 452, confirming the [OIII] luminosity to be affected by residual extinction in presence of mild absorption, especially for ''buried'' AGN such as 3C 452.
Three of our targets are powerful FRII radio galaxies, which is shown by their high luminosity and absorption; this makes them the most luminous and absorbed AGN of the BAT Seyfert survey despite the inversely proportional N$_{\rm H}$ - L$_{\rm X}$ relation.
}
   {}


   \maketitle
%

\section{Introduction}
Absorption by circumnuclear gas and dust surrounding the accreting supermassive black hole (SMBH) is a common feature of active galactic nuclei (AGN). 
According to the unified model (\citealt{ant93}), different observational features can be explained on the basis of the AGN orientation with respect to the observer, such that an object is classified as unobscured Type I AGN if the central broad line region appears in the optical spectrum, while the lack of these features occurs if the nuclear regions are seen at larger inclination angles and the line of sight intercepts optically thick cold matter distributed in a toroidal geometry.
This model is supported by both X-ray (\citealt{bas99}, \citealt{toz06}, \citealt{cap06}) and infrared (IR) observations (\citealt{alo06}, \citealt{fio08}), where a high percentage of AGN have X-ray column densities greater than $3\times10^{22}$ cm$^{-2}$.
\\
If the bulk of the SMBH growth is expected during an obscured phase of the AGN evolution (\citealt{hop06}), and its accretion is tightly connected to the galaxy bulge evolution and starburst activity in a feedback process (\citealt{fer00}, \citealt{bra05}), the complete census of the obscured AGN is a fundamental key to unlock the SMBH growth history.
In the local Universe, a significant fraction of Seyfert 2 (\citealt{1999ApJ...522..157R}, \citealt{gua05}) are Compton-thick (N$_{\rm H}>1.5\times10^{24}$ cm$^{-2}$). Current synthesis models of the cosmic X-ray background (CXB), given by the contribution of unobscured and obscured AGN integrated over the cosmic time (\citealt{set89}, \citealt{com95}), predict that a high percentage of Compton-thick AGN contribute to the energy density peak at 30 keV (\citealt{gil01}).
\\ 
Despite the high penetrating power of X-ray photons, the nature of this population is still elusive because of the strong obstacles against their detection, because for column densities above $3\times10^{23}$ cm$^{-2}$ the X-ray flux from the nucleus is reduced by more than a factor of 10.
In this context, hard X-ray surveys above 10--15 keV performed in recent years by Swift/BAT (14--195 keV, \citealt{tue10}) and INTEGRAL (10--100 keV, \citealt{bec09}) represent a unique tool for obtaining a complete, unbiased sample of AGN for column densities $<3\times10^{25}$ cm$^{-2}$. The need for X-ray selected AGN surveys is also of particular interest if we consider that more than 50\% of the X-ray AGN are not detected at other wavelengths and probably trace different AGN populations and/or evolutionary stages (\citealt{2009ApJ...696..891H}).
The discovery of a new class of ``buried'' AGN (\citealt{ued07}) from the BAT hard X-ray survey is a direct consequence of the new observing window. 
\\
The full characterization of the obscuring matter surrounding the SMBH requires more than the N$_{\rm H}$ evaluation. Other important observational clues are found along the entire X-ray spectrum: from the large equivalent widths of the Fe K line to the flattened slope and the broad bump peaking around 20--30 keV due to Compton down-scattering of high-energy photons.
These additional diagnostics for obscuration are necessary to avoid an underprediction of the column density that could affect the computed X-ray luminosity even in the hard E$>10$ keV band (\citealt{2011lamassa}, \citealt{2008melendez}).
\\
The Hard X-ray Detector (HXD) on board the NASA/JAXA Suzaku mission extends the bandpass of the X-ray focusing telescope up to 600 keV with an unprecedent sensitivity in the 10--100 keV energy range thanks to the low in-orbit background (\citealt{tak07}). 
\\
The Swift/BAT nine-month survey observed 153 AGN, all with ultra-hard X-ray BAT fluxes (14-195 keV) in excess of $10^{-11}$ erg cm$^{-2}$ s$^{-1}$ and an average redshift of 0.03. In this paper we analyze the follow-up Suzaku observations of four of the most luminous BAT AGN ($44.73 < \rm{Log\;L}_{\rm BAT} < 45.31$).
\\
The reason for the choice to focus on absorption and reflection of high X-ray luminosity AGN is the dependency of the obscured fraction on the AGN luminosity. The difficulty in mapping the space density of highly obscured AGN in the local Universe is overcome by the large uncertainties in determining the column density distribution as a function of luminosity and cosmic time, which is strictly correlated to the accretion history and evolution of AGN. While there is still debate on the redshift distribution of obscured AGN (\citealt{laf05}, \citealt{bal06}, \citealt{tre10}), in the past decade X-ray (\citealt{ued03}, \citealt{laf05}) optical (\citealt{sim05}) and infrared (\citealt{fio09}) studies seem to confirm the inverse proportionality of the obscured population as a function of the luminosity, with a moderately obscured to unobscured sources ratio decreasing from 4 to 1 for L$_{\rm X}\leq10^{42}$ and L$_{\rm X}\geq10^{45}$, respectively (\citealt{gil07}). In this context, characterizing the obscuring matter surrounding high-luminosity AGN helps to separate the various interpretations (\citealt{has08}, \citealt{dia09}) and to better constrain the N$_{\rm H}$-L$_{\rm X}$ correlation. 
\\
In this paper, the extended Suzaku bandpass is exploited to study within the same observation the soft/hard X-ray emission of four AGN from the nine-month Swift/BAT survey (\citealt{tue08}). The column density, scattered/reflected emission, and the properties of the Fe K line are fully analyzed along with their variability, while the nine-month-averaged Swift/BAT spectrum helps to better constrain the spectral shape in the hard X-ray domain.

%
\begin{table*}
\caption{\label{tab:targ}}             
\begin{minipage}{\textwidth}
\centering          
\renewcommand{\footnoterule}{}  
\begin{tabular}{c|cccccccc }    
\multicolumn{9}{c}{\textsc{List of Targets}\footnote{The position and redshift (J2000 coordinates) are taken from the NASA/IPAC Extragalactic Database.}} \\
\hline\hline       
\multirow{2}{*}{Source} & \multirow{2}{*}{R.A.} & \multirow{2}{*}{Dec.}  & \multirow{2}{*}{Redshift} &  \multirow{2}{*}{Log L$_{\rm BAT}$\footnote{Logarithm of the luminosity in the 14--195 keV band in erg s$^{-1}$ from the BAT nine-month AGN survey.}}& \multirow{2}{*}{Optical type\footnote{FRII RG = Fanaroff-Riley Class II Radio Galaxy; GRG = Giant Radio Galaxy.}} & \multirow{2}{*}{Radio type\footnote{GRG = Giant Radio Galaxy; Sy = Seyfert; QSO = Quasar; / = No radio counterpart is associated to the source.}} & \multirow{2}{*}{Host Galaxy\footnote{The AGN host galaxy is indicated as E (elliptical), ? (optical image too faint to be classified), Merg? (possible galaxy merging).}} & \multirow{2}{*}{Ref.\footnote{1 = \cite{lai94}, 2 = \cite{jack97}, 3 = \cite{bla92}, 4 = \cite{tue08}, 5 = \cite{scho98}.}}\\
 & & & & & & & & \\
\hline                    
\multirow{2}{*}{3C 452} & \multirow{2}{*}{341.4532} & \multirow{2}{*}{39.6877} & \multirow{2}{*}{0.0811} & \multirow{2}{*}{44.73} & \multirow{2}{*}{Sy 2} & \multirow{2}{*}{FRII RG} & \multirow{2}{*}{E} & \multirow{2}{*}{1, 2, 3}\\
 & & & & & & & & \\
\multirow{2}{*}{3C 105} & \multirow{2}{*}{61.8186} & \multirow{2}{*}{3.7071} & \multirow{2}{*}{0.089} & \multirow{2}{*}{44.83}  & \multirow{2}{*}{Sy 2} & \multirow{2}{*}{FRII RG} & \multirow{2}{*}{?} & \multirow{2}{*}{1, 2, 3}\\
 & & & & & & & & \\
\multirow{2}{*}{J0318.7+6828}  &  \multirow{2}{*}{49.5791} & \multirow{2}{*}{68.4921} & \multirow{2}{*}{0.0901} & \multirow{2}{*}{44.85} & \multirow{2}{*}{Sy1.9} & \multirow{2}{*}{FRII GRG} & \multirow{2}{*}{?} &\multirow{2}{*}{4, 5}\\
 & & & & & & & & \\
\multirow{2}{*}{J0918.5+0425}  & \multirow{2}{*}{139.5011} & \multirow{2}{*}{4.4184} & \multirow{2}{*}{0.1564} & \multirow{2}{*}{45.31} & \multirow{2}{*}{QSO 2} & \multirow{2}{*}{/} & \multirow{2}{*}{Merg?} &  \multirow{2}{*}{4}\\
 & & & & & & & & \\
\hline                  
\end{tabular}
\end{minipage}
\end{table*}
\section{Observations}
The targets of our analysis are Swift J2246.0+3941 (3C 452), J0407.4+0339 (3C 105), J0318.7+6828, and J0918.5+0425. We summarize their basic parameters in Table \ref{tab:targ} and also list the host galaxy, AGN radio, and optical classification collected from the literature.
The first two targets, J2246.0+3941 and  J0407.4+0339, are well-known 3C catalog FRII Radio Galaxies, and we refer to them as 3C 452 and 3C 105, while J0318.7+6828 is an FRII Giant Radio Galaxy (GRG) with a linear size of 2 Mpc.
Optical spectra, available for all the sources, identify our targets as Type II AGN. 
The last, and farthest, J0918.5+0425 has been identified as an AGN for the first time by the nine-month Swift/BAT survey, but was then not detected in the 4.8$\sigma$ cut of the 22-month survey, most likely because of variability (\citealt{tue10}).
\\
The Japanese Suzaku (ASTRO-E2) X-ray mission, launched in 2005, achieves its wide (0.2--600 keV) X-ray energy range through combining four imaging CCD cameras (the X-ray imaging spectrometers, XIS) with a non-imaging, collimated hard X-ray detector (HXD).
Each XIS detector, three front- (XIS-0,2,3) and one back-illuminated (XIS-1), is located in the focal plane of a dedicated X-ray telescope with a bandpass in the range 0.2--12 keV, while the HXD consists of Si PIN photo-diodes and GSO scintillation counters, covering the 10--70 keV and 40--600 keV energy range (\citealt{mit07}).
\\
Suzaku observed our sources (in HXD nominal pointing) between June 2007 and February 2008 (see observation details in Table \ref{tab:su_log}). Since the FI (front-illuminated) XIS-2 detector became inoperable in 2006 November, no XIS-2 data are available. 
We present the analysis of XISs and HXD/PIN data because the flux of our targets was too faint to be detected by HXD/GSO.
In addition to the Suzaku data, we also integrate the 14--195 keV BAT spectra in the spectral analysis, which were time-averaged over the nine-month survey to better constrain the fit.
\\
Except for 3C 105, which is characterized by shorter detections, the net (dead-time corrected) exposure spans from about 60 to 80 ks for the XIS-BI, and from about 50 to 65 ks for the high-energy detector.
\begin{table*}
\caption{ \label{tab:su_log}  }
\begin{minipage}{\textwidth}
\centering
\renewcommand{\footnoterule}{}  
\begin{tabular}{c|cccc}
\multicolumn{5}{c}{\textsc{Suzaku Observation Log}}\\
\hline
\hline
\multirow{2}{*}{Source} & \multirow{2}{*}{Obs ID} & \multirow{2}{*}{Date} & \multirow{2}{*}{Exposure\footnote{Exposure time (s) and background-subtracted count rate (cts s$^{-1}$) for XIS-BI (back-illuminated) and HXD/PIN respectively in the 0.5--10 keV and 10--30 keV band}}  & \multirow{2}{*}{Ct. Rate$^{a}$}
  \\
 & & & & \\
\hline
\multirow{2}{*}{3C 452} & \multirow{2}{*}{702073010} & \multirow{2}{*}{2007-06-16} & \multirow{2}{*}{66696, 57383} &  \multirow{2}{*}{0.037, 0.039}\\
 & & & & \\
\multirow{2}{*}{3C 105} & \multirow{2}{*}{702074010} & \multirow{2}{*}{2008-02-05} &  \multirow{2}{*}{38319, 45518} &  \multirow{2}{*}{0.031, 0.024} \\
 & & & & \\
\multirow{2}{*}{J0318.7+6828} & \multirow{2}{*}{702075010} & \multirow{2}{*}{2007-09-22} &  \multirow{2}{*}{77140, 65631} &  \multirow{2}{*}{0.091, 0.027}\\
 & & & & \\
\multirow{2}{*}{J0918.5+0425} &\multirow{2}{*}{702076010} & \multirow{2}{*}{2007-11-04} &  \multirow{2}{*}{61019, 51899} &  \multirow{2}{*}{0.024, 0.006}\\
 & & & & \\
\hline
\end{tabular}
\end{minipage}
\end{table*}%

\section{Data analysis}
Observation products (spectra, images, light curves) were extracted from the cleaned version 2.3.12.25 processed event files distributed by the Suzaku team, with the most recent \texttt{HEAsoft} package. The XIS source spectra and light curves, generated by combining the $3\times3$ and $5\times5$ edit modes in \texttt{xselect}, were extracted from circular regions with a radius ranging from $2.2'$ to $4.2'$. The minimum $2.2'$ radius region was used for 3C 452, allowing us to take more than 80\% of the point source photons (see Suzaku technical description) while avoiding contamination from external sources in the field. A $4.2'$ radius region, which collects 99\% of the point source flux, was used for the other sources because there are no apparent confusing sources (see Sec. \ref{sec:pmc}) inside the extracting region.
Background spectra were extracted from nearby free-emission regions of about $2.2'$--$5'$ radius. 
After producing the rmf and arf response matrices with the \texttt{xisrmfgen} and \texttt{xissimarfgen} tools provided by the Suzaku team, we combined the two XIS-FI spectra and responses with \texttt{addascaspec} for faster spectral fits, grouping (with \texttt{grppha}) both XIS-FI and XIS-BI source spectra using a factor greater than 25 cts/bin.
\\
Because HXD/PIN is a non-imaging detector, a model of the time-variable non X-ray background (NXB) is provided by the HXD team (dead-time corrected) with a systematic uncertainty of about 1.3\% at 1$\sigma$ confidence level in the 15--40 keV band (for a 10 ks exposure, see \cite{miz08}). Since the NXB event rate is ten times higher than the real background, the particle background spectra and light curves were increased by a factor of 10 with the \texttt{fparkey} and \texttt{fcalc} tools to suppress the Poisson error.
A common good time interval (GTI) between the PIN cleaned event file and the NXB was created to obtain dead-time corrected spectra and light curves.
Like the NXB, the PIN also still suffers from CXB contamination (about 5\%). This was estimated from the PIN response file for the flat emission distribution (provided by the HXD team) and then simulated with \texttt{xspec} using the HEAO-1 CXB spectrum (\citealt{bol87}). The particle background and CXB spectra were combined using \texttt{mathpha}. The PIN response files, available from the Suzaku CALDB as generated in June 2009, were chosen according to the HXD nominal pointing and the observation epoch.

\section{Spectral analysis}\label{spectral_analysis}
Using the X-ray spectral fitting package \texttt{Xspec v12} (\citealt{xspec}), three main models were tested to describe the primary emission from the AGN. Each target was initially fitted with the simplest case; if the model was unable to describe the spectrum or the fit was significantly improved by introducing additional components, we adopted more complex models. In addition to the continuum emission, a Gaussian profile was added to characterize the Fe K$\alpha$ fluorescence line at 6.4 keV (\texttt{\textbf{zgauss}}) as the result of reprocessing in the surrounding material (e.g., accretion disk, torus, ionized gas). The physical width was kept free to vary. If the spectrum showed no significant fluorescence line, we derived an upper limit of the equivalent width, fixing the line centroid (in the source rest frame) at 6.4 keV.
\\
We used the \texttt{\textbf{zphabs}} code to model the photoelectric absorption by cold matter across the line of sight with cross sections from \cite{cross}. The neutral absorption taking place in our galaxy was accounted for (even when not explicitly mentioned) with the \texttt{\textbf{tbabs}} model (\citealt{tbabs}). The galactic column density, from the HI map of \cite{nh90}, was provided by the \texttt{nh} program of the \texttt{HEAsoft} package and is listed in Table \ref{tab:fit_res}. Solar abundances as given by \cite{abund} were assumed throughout the analysis, with cosmological parameters of H$_{0}$ = 70 km s$^{-1}$ Mpc$^{-1}$, $\Omega_{\rm M} = 0.27$ and $\Omega_{\rm vac} = 0.73$.

\subsection{Simple model}
The X-ray continuum emission from a Type II AGN is characterized, as a first-order approximation, by an absorbed power law (\citealt{mus93}), resulting from thermal Comptonization of soft-seed accretion-disk photons by a hot diffuse corona (\citealt{hm91}). Defining as $\rm N_{\rm H}$ and $\sigma(\rm E)$ the column density and photoelectric cross section respectively of the cold matter surrounding the black hole, this simple model takes the form F(E) = exp$\{- \rm N_{\rm H}\sigma(\rm E)\}$ E$^{-\Gamma}$ ($\Gamma$ represents the photon index), translated into \texttt{\textbf{zphabs*zpowerlaw}} in Xspec terminology.
\\
We assumed the absorber to be neutral, although ionized absorbers in radio loud AGN have recently been found (e.g., in the BLRG 3C 445, \cite{bra11}), because the data quality does not allow such a detailed analysis.

\subsection{Partial-covering model}\label{sec:pmc}
Several Type II AGN spectra show, in addition to the absorbed primary emission, a secondary less absorbed, or not at all absorbed, emission at lower energies (\citealt{tur97}). Many interpretations could be applied to this observational feature: it could be 1) the result of contamination by other sources (e.g., X-ray binaries/starburst emission, \cite{mai98}); or 2) a clumpy, dusty absorber that partially covers the nucleus (\citealt{mal99}); or 3) AGN light scattered back into our line of sight by hot gas (\citealt{mat96}); or 4) the sum of unresolved emission lines from photoionized gas (\citealt{bia06}).
\\
Given the point spread function (PSF) of the XIS instruments, we cannot rule out other sources on the basis of the Suzaku imaging capabilities alone. For this reason, both the Swift/XRT and, when available, the XMM-Newton and Chandra images of our targets were analyzed to investigate contamination by point sources or the lobes of the associated radio galaxy.
\\ 
If we interpret the secondary component as intrinsic emission from the AGN, this complex spectrum can be described by the partial-covering model (see \citealt{holt80}), where a fraction of the primary emission escapes without being absorbed. It requires the hydrogen column density and the covering fraction of the partial absorber (f$_{\rm c}$), and can be written as F(E) = \{f$_{\rm c}$ exp$\{- \rm N_{\rm H}\sigma(\rm E)\}$ + (1 - f$_{\rm c}$)\} E$^{-\Gamma}$.
This model is useful to describe not only a clumpy absorber but also scattered nuclear emission escaping from different regions, characterized by lower column density, and redirected to our line of sight, assuming the same slope for the scattered power law. We refer to the value 1 - f$_{\rm c}$ as the scattering fraction f$_{\rm scatt}$ of the absorbed power law. The partial-covering code takes the form \texttt{\textbf{zpcfabs*zpowerlaw}}. If f$_{\rm c}<1$, a fraction (1 - f$_{\rm c}$) of the primary emission escapes without being absorbed, while a f$_{\rm c}=1$ returns the simple absorbed model.
\\
The soft X-ray emission could also be thermal, due to collisionally ionized diffuse gas that could be present in the host galaxy or nearby the AGN itself \citep{iso02}, or the sum of unresolved lines produced by photoionized gas (\citealt{ree10}). The XIS spectral resolution does not allow us to separate the two sources, and we only searched for additional thermal gas with the \texttt{\textbf{apec}} component to better constrain the scattered/partially covered power law.

\subsection{Double power law model}
In general, the second, scattered power-law could be absorbed by less dense clouds and be characterized by a different slope than the direct emission. The third model increases the complexity of the fit by adopting two different absorbed power-laws of the form F(E) = exp$\{- \rm N_{\rm H}^{\rm a}\sigma(\rm E)\}$ E$^{-\Gamma^{\rm a}}$ + exp$\{- \rm N_{\rm H}^{\rm b}\sigma(\rm E)\}$ E$^{-\Gamma^{\rm b}}$, where a and b mark distinct values, or \texttt{\textbf{zphabs*zpowerlaw + zphabs*zpowerlaw}} in Xspec.

\subsection{Compton reflection}
Given the extended energy range of the PIN, the additional contribution of Compton reflection by optically thick matter could arise in the hard part of the spectrum.
The Xspec \texttt{\textbf{pexrav}} code models the angle-dependent Compton reflection of an incident exponentially cut-off power law by a plane-parallel, semi-infinite medium of cold electrons that approximate an accretion disk (\citealt{pexrav}). We applied this code to a more general case where the AGN primary radiation could be reflected by optically thick matter present in the accretion disk or in the surrounding torus/absorber (see the caveats in \cite{mur09}), resulting as the sum of the incident cut-off power-law and the reflection component. The amount of reflection depends on both geometry and composition of the reflector, parameterized in the model by the inclination of the reflecting material and the gas abundance, with the possibility of varying the Fe abundance alone. The model also returns the relative strength R (rel$_{\rm refl}$ in the code) of the reflection component to that of the incident cut-off power law component, defined as $\rm R\equiv\Omega/2\pi$, where $\Omega$ is the solid angle covered by the reflector. If $\rm R=0$, the pexrav model translates into a simple exponentially cut-off power law.
Each model (simple model, partial-covering model, double power-law model) was tested with and without the \texttt{\textbf{zphabs*pexrav}} code for all sources. When a significant reflection component was not found from the spectral fitting, we evaluated an upper limit to the relative reflection by fitting the spectra with a partially covered pexrav model of the form \texttt{\textbf{zpcfabs*pexrav}}. As previously noted, for $\rm R=0$ this model turns into a partial-covering model.

\section{3C 452}\label{sec:452}
The source 3C 452 is an FRII radio galaxy with a symmetrical double-lobe morphology and a total angular extent of about $1'\times4'$ (\citealt{bla92}).
Because it does not lie in a rich cluster, it is not contaminated by thermal X-ray emission related to the intracluster medium (\citealt{iso02}, hereafter I02). From the Hubble observation, no UV excess (\citealt{wills02}) and no optical central compact core are detected (\citealt{chi02}), while the NICMOS near-IR instrument reveals a faint compact source 1" southwest of the nucleus (\citealt{mad06}), confirmed by the weak mid-IR core detected by Spitzer (\citealt{ogle06}).
\\
The Chandra/ACIS instrument observed the source on August 2001: a detailed spectral analysis of the two lobes and central AGN emission is reported by I02, and the results for the AGN properties are confirmed by \cite{eva06}. From I02, the soft X-ray emission from the extended radio structure is fitted by a power law, interpreted as inverse compton (IC) and significant in the 2--5 keV band, with the addition of the Raymond-Smith (RS) thermal emission in the $0.5-2$ keV band from hot, diffuse gas associated to the host galaxy. 
\cite{hk10} confirmed this emission throughout the region covered by the radio lobes, although with the non-thermal emission dominant near the hot spots, but they associated the thermal emission to the radio galaxy cocoon or a bounding shock and not to the actual intergalactic medium.
   \begin{figure}[h!]
   \centering
   \includegraphics[width=.48\textwidth, angle=0]{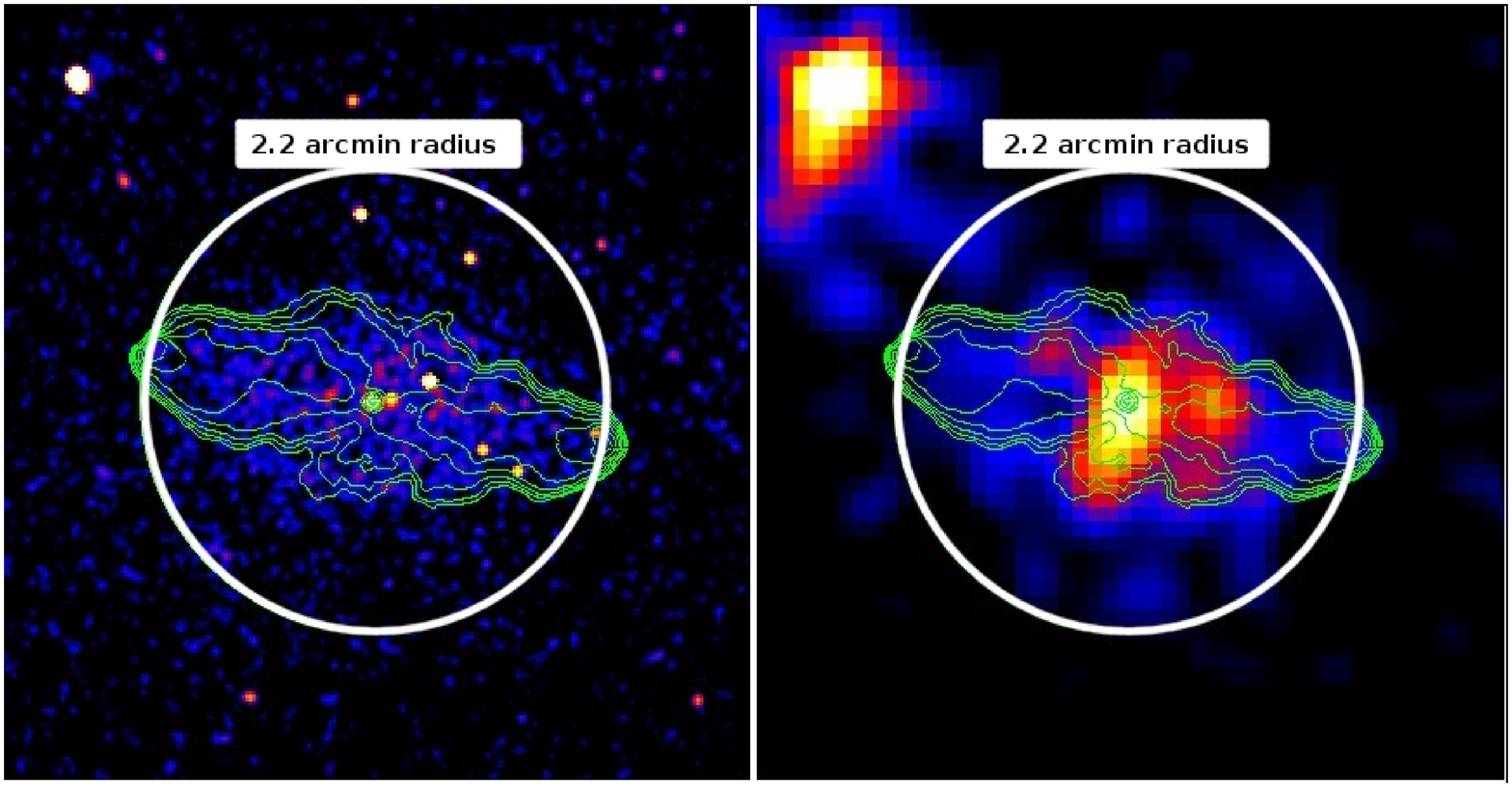}\\
   \includegraphics[width=.48\textwidth, angle=0]{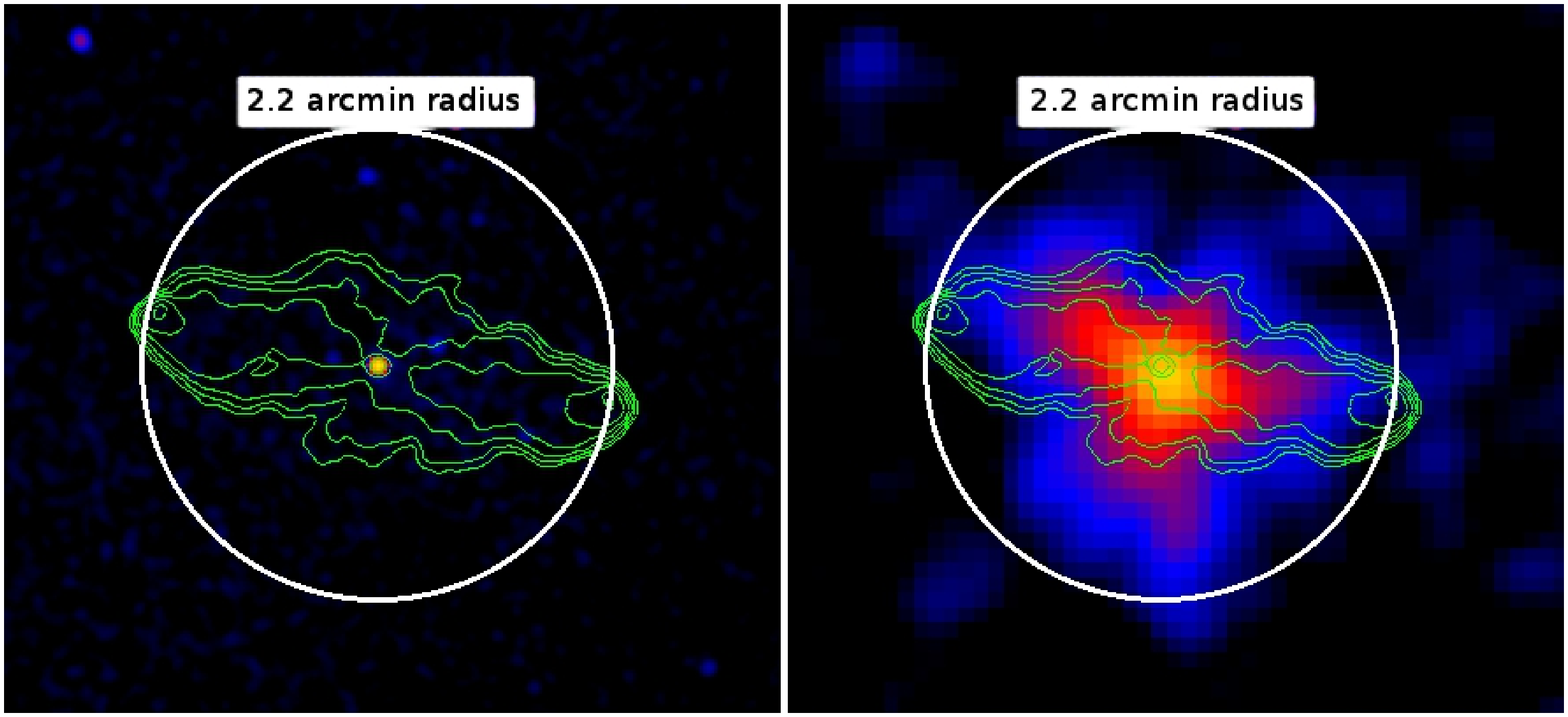}   
   \caption{Chandra/ACIS (left panel) and Suzaku/XIS1 (right panel) images of 3C 452 in the 0.5--2 keV (top panels) and 2--10 keV (bottom panels) energy ranges. The 1.4 GHz VLA radio map is superimposed with green contours and the white circle indicates the Suzaku $2.2'$ radius extracting region.}
              \label{fig:acis_3c452_fig}%
    \end{figure}
\subsection{Extracting the AGN emission}
\begin{figure}[h!]
   \centering
   \includegraphics[width=.25\textwidth, angle=-90]{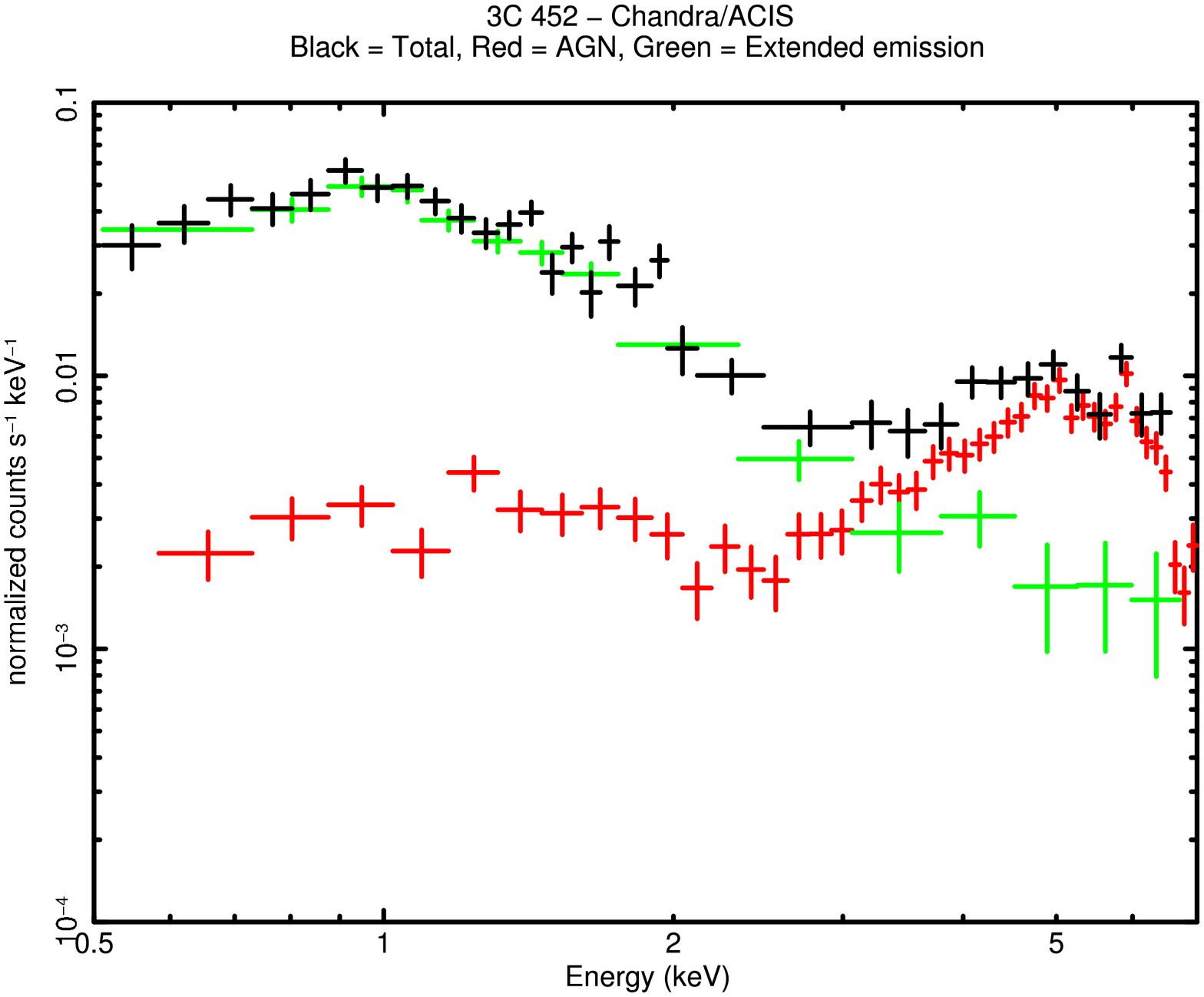}\\
   \vspace{1cm}
   \includegraphics[width=.25\textwidth, angle=-90]{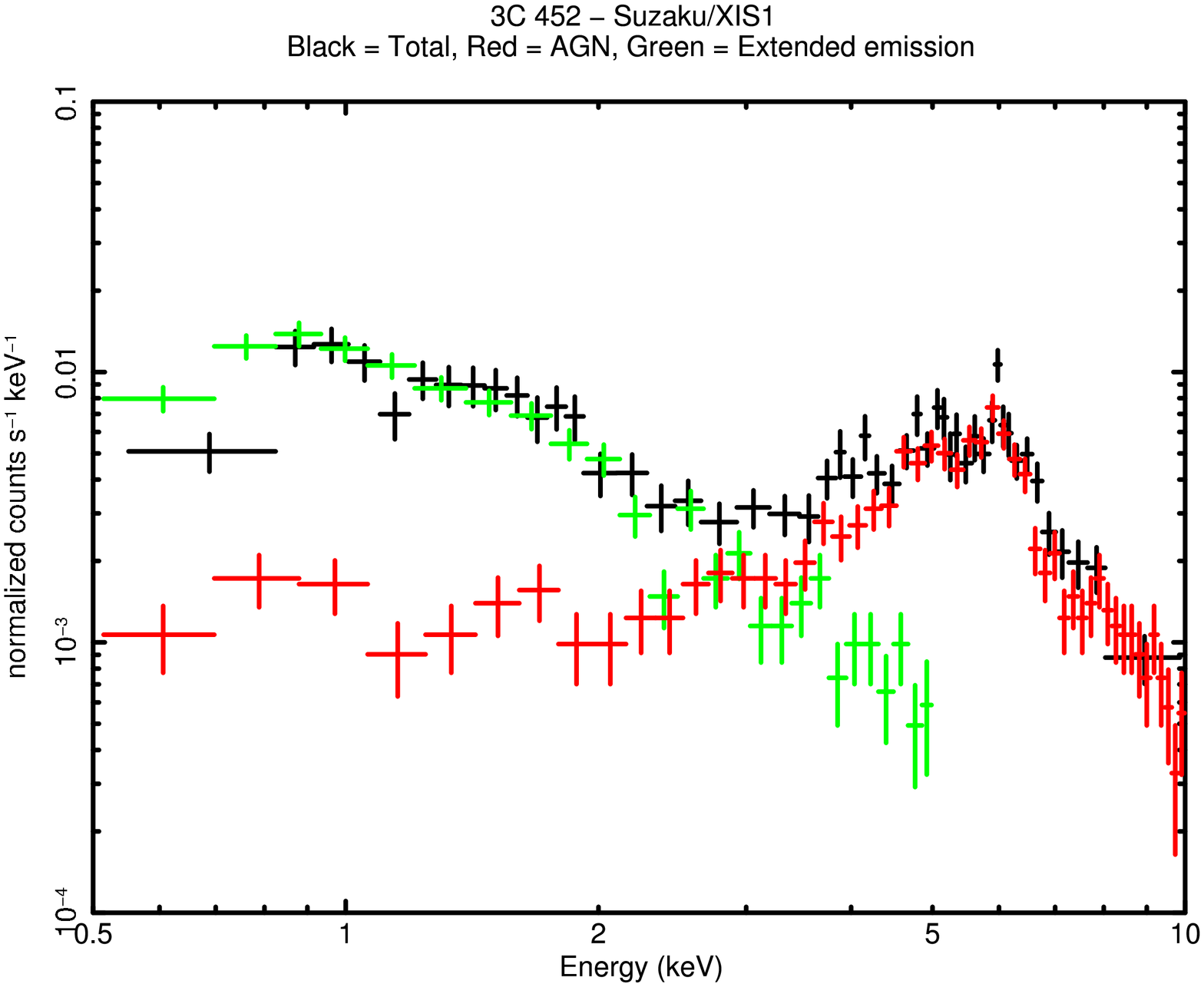}
   \caption{Top panel: Chandra/ACIS spectrum of the $2.2'$ radius region centered on the AGN (black line), the AGN alone (red line), and the extended region after subtracting the AGN core (green line). Bottom panel: the Suzaku/XIS1 observed spectrum is shown in black, while the red and green lines refer to the XIS1 simulated spectra of the AGN and the extended emission.}
              \label{fig:sim_fig}%
    \end{figure}

In the analysis of the Suzaku/XIS image, the required extracting region of the central point source covers part of the extended radio galaxy because the PSF is wider than that of Chandra/ACIS: the AGN emission is contaminated by the diffuse emission at energies lower than 5 keV. This can be clearly seen in the top panels of Figure \ref{fig:acis_3c452_fig}, which shows the Chandra/ACIS and XIS1 0.5--2 keV smoothed image of 3C 452 in the left and right panels, while the white circle refers to the $2.2'$ radius extracting region. The 1.4 GHz Very Large Array (VLA) radio map (Laing R.A. unpublished) is superimposed with green contours: the soft, cocoon shaped X-ray emission traces the radio contours in the ACIS image while it is not visible in the 2--10 keV energy selection. 
\\
Because we aim to analyze the AGN spectrum, it is necessary to separate the X-ray photons of the central core from the extended emission. The lobe-induced contamination in the XIS spectrum was checked by simulating the diffuse and core detection of the XIS1 on the basis of the ACIS observation with the \texttt{\textbf{xissim}} tool.
We extracted the ACIS total spectrum from the same region on which the Suzaku 3C 452 analysis is based: the central core counts were taken from a 6" circular region, while the extended emission was obtained from subtracting the core region from the total one (Figure \ref{fig:sim_fig}, top panel). 
The two Chandra spectra were fitted according to the models and best-fit parameters reported in I02, leaving the normalization free to vary. Applying the XIS1 response matrix to the resulting AGN and extended emission spectral models, we obtained a count rate of a fake XIS1 detection of the two components. 
\\
The bottom panel of Figure \ref{fig:sim_fig} shows the simulated XIS1 detection (the AGN emission in red and the extended emission in green) and, as a comparison, the real observation (in black), which is the sum of the diffuse and central core emission. The simulation reproduces the real detection well, i.e., the Suzaku/XIS spectra are contaminated by the diffuse emission below 3 keV. To model the AGN emission, two methods are possible: 1) the XIS spectrum is analyzed only above 3 keV, losing all information about the possible scattered/partially covered component; 2) the model of the soft X-ray emission, based on the I02 results, is added to the AGN model, and the full energy band XIS spectrum (0.5--10 keV) is analyzed. We selected this last option to be able to also explore the possible partially-covered emission.

\subsection{Spectral fitting}
We analyzed the spectrum in the full energy band by adding the XIS (FI and BI in the 0.5--10 keV energy range), the PIN (10--30 keV), and the nine-month-averaged BAT (14--195 keV) data. The normalization between XIS and PIN flux was accounted for using a 1.18 factor in the PIN model, given by the cross-correlation of the Crab observation in HXD nominal pointing, while the BAT normalization was kept free. Because the extended emission is not expected to vary during the six year time interval between the Chandra and Suzaku observations, we can model the extended emission on the basis of the Chandra spectral parameters: \texttt{\textbf{zpowerlaw + raymond}} (the thermal emission fitting is based on the calculations of \cite{ray77}), with a power-law photon index of 1.68 and a gas temperature of 1.36 keV, as reported in I02.
The relative flux for the extended emission derived from spectral fitting between ACIS and Suzaku agrees within the cross-calibration error of the two instruments.
The simple absorbed power law leaves an excess in the PIN spectrum, and the reflection component is needed. In the \texttt{\textbf{pexrav}} model we could not determine the high-energy cut-off, which was fixed at 200 keV, or the inclination and chemical composition of the reflector, fixed at 63$^{\circ}$ (the default value) and solar abundance. The partial-covering model did not improve the fit, given the model-crowded soft band, but it allowed to fix an upper limit to the scattering fraction, which is below 0.5\%. A potential absorption of the secondary power law could not be constrained and was removed from the model. The best-fit model is \texttt{\textbf{zpcfabs*zpowerlaw + zgauss + pexrav}}, and the main parameters are reported in Table \ref{tab:fit_res}. 
   \begin{figure}[h!]
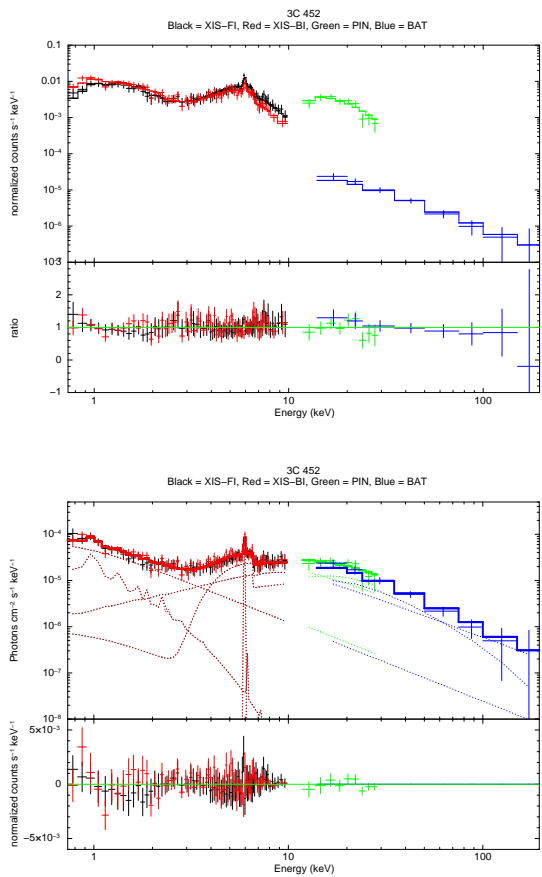

   \centering
   \includegraphics[width=.3\textwidth, angle=-90]{pap_3c452_xis_pin_allbat_spectrum.ps}\\
   \vspace{0.5cm}   
   \includegraphics[width=.3\textwidth, angle=-90]{pap_3c352_pcm_defA_allbat_unf_RES.eps}
   \caption{Folded (top panel) and unfolded (bottom panel) spectra and best-fit models of 3C 452: black = XIS-FI, red = XIS-BI, green = HXD/PIN, blue = Swift/BAT.}
              \label{fig:3c452_fig}%
    \end{figure}
\subsection{Fitting results}
The source 3C 452 is highly absorbed, with a local column density N$_{\rm H}$ of $4\times10^{23}$ cm$^{-2}$. The hard energy band of the spectrum is dominated by Compton reflection of the intrinsic power-law photons onto a cold, thick matter, and the source seems to see the absorber covering a solid angle greater than 4$\pi$: the relative strength of the reflection component to the transmitted one, R, is higher than 400. This value is unphysical, but it can be mathematically explained as the result of an inhomogeneous absorber, with a significant fraction of the solid angle covered by a gas thicker than that along the line of sight (\citealt{2002ApJ...571..234R}). The nuclear emission is flattened, with a power-law photon index of 1.55, while the average Seyfert II slope ranges from 1.7 to 1.9 (\citealt{tur89}, \citealt{na94}). 
The relative normalization of the BAT spectrum is close to 1: the lack of variability is consistent with 3C 452 behaving as a Compton-thick AGN (see Section \ref{sec:var}). However, the Fe K$\alpha$ line, centered at 6.43 keV (in the source rest frame) and narrow, shows an equivalent width (EW) of only 164 eV, in contrast with what is expected from Compton-thick sources (see Sec. \ref{sec:disc}). 

\begin{table*}
\caption{ \label{tab:fit_res}  }
\begin{minipage}{\textwidth}
\centering
\renewcommand{\footnoterule}{}
\begin{tabular}{c|ccccccccc}
\multicolumn{10}{c}{\textsc{Suzaku Spectral Fit Compared Results}}\\
\hline
\hline
 &  &  &  &  &  &  &  &  &\\
	\multirow{2}{*}{Source} & n$_{\rm H}$\footnote{Galactic column density provided by the \texttt{nh} tool, using the HI map by \cite{nh90}.} & N$_{\rm H}$\footnote{Source intrinsic column density.}& \multirow{2}{*}{$\Gamma$} & \multirow{2}{*}{f$_{\rm scatt}$\footnote{Fraction of the scattered component relative to the intrinsic power law.}} & Fe K E\footnote{Centroid energy and equivalent width respect to the whole continuum of the Fe fluorescence K$\alpha$ line at the rest frame of the source redshift.} &  Fe K EW$^{d}$ &  \multirow{2}{*}{R\footnote{The relative strength of the reflection component to the transmitted one, defined as R$=\Omega/2\pi$, where $\Omega$ is the solid angle of the reflector viewed from the nucleus.}} & BAT & \multirow{2}{*}{$\chi^{2}/\rm d.o.f$} \\
& [$10^{22}$ cm$^{-2}$] & [$10^{22}$ cm$^{-2}$] & & & [keV] & [eV] & & Norm.\footnote{Normalization factor of the BAT flux respect to the XIS detection.} & \\
 &  &  &  &  &  &  &  &  &\\
\hline
\multirow{2}{*}{3C 452} & \multirow{2}{*}{0.12} & \multirow{2}{*}{${43.52}^{+10.85}_{-6.92}$} & \multirow{2}{*}{${1.55}^{+0.14}_{-0.11}$} & \multirow{2}{*}{$< 0.5\%$} & \multirow{2}{*}{${6.43}^{+0.03}_{-0.03}$} & \multirow{2}{*}{${164}^{+44}_{-45}$} & \multirow{2}{*}{$>400$} & \multirow{2}{*}{${0.90}^{+0.25}_{-0.15}$} & \multirow{2}{*}{116.76/175}\\
 &  &  &  &  &  &  &  &  &\\
\multirow{2}{*}{3C 105} & \multirow{2}{*}{0.12} & \multirow{2}{*}{${45.96}^{+6.24}_{-6.56}$} & \multirow{2}{*}{${1.78}^{+0.20}_{-0.19}$} & \multirow{2}{*}{${1.4^{+0.9}_{-0.5}\%}$} & \multirow{2}{*}{${6.40^{+0.07}_{-.0.05}}$} & \multirow{2}{*}{${136}^{+75}_{-62}$} & \multirow{2}{*}{$<1.8$} & \multirow{2}{*}{${1.68}^{+0.61}_{-0.45}$} & \multirow{2}{*}{68.18/97}\\
 &  &  &  &  &  &  &  &  &\\
\multirow{2}{*}{J0318.7+6828} & \multirow{2}{*}{0.35} & \multirow{2}{*}{${5.26}^{+0.42}_{-0.41}$} & \multirow{2}{*}{${1.55}^{+0.08}_{-0.08}$} & \multirow{2}{*}{${1.2}^{+0.7}_{-0.8}\%$} & \multirow{2}{*}{6.4(f)} & \multirow{2}{*}{${63}^{+32}_{-27}$} & \multirow{2}{*}{$<1.4$} & \multirow{2}{*}{${1.08}^{+0.27}_{-0.23}$} & \multirow{2}{*}{167.68/172} \\
 &  &  &  &  &  &  &  &  &\\
\multirow{2}{*}{J0918.5+0425} & \multirow{2}{*}{0.035} & \multirow{2}{*}{${16.33}^{2.27}_{-2.13}$} & \multirow{2}{*}{${1.72}^{+0.20}_{-0.19}$} & \multirow{2}{*}{${0.9}^{+0.6}_{-0.4}\%$} & \multirow{2}{*}{6.4(f)} &  \multirow{2}{*}{$<98$} & \multirow{2}{*}{$<2.6$} & \multirow{2}{*}{${1.74}^{+1.16}_{-0.87}$} & \multirow{2}{*}{79.21/132}\\
 &  &  &  &  &  &  &  &  &\\
\hline
\end{tabular}
\end{minipage}
\end{table*}%

\section{3C 105}\label{105}
The source 3C 105 is an FRII narrow-line radio galaxy with a spatial size of the radio emission of 764 kpc (\citealt{baum88}, \citealt{hard98}), associated to an elliptical host galaxy (\citealt{do07}). From the radio map of \cite{lea97}, both jets are detected, but the radio emission is dominated by the southern hotspot and the wealth of other knots and structures suggests (\citealt{lea97}) that the system is rapidly evolving, with a short lifetime. Chandra detected X-ray emission from the core and southern hot spot of 3C 105 in December 2007 (\citealt{mas10}, hereafter M10), both from the region where the jet appears to enter the hot spot and at the terminal hotspot itself. No extended, thermal X-ray emission is detected. A faint, unresolved near-IR source is visible $4''$ to the west-southwest off the nucleus (\citealt{mad06}), while we were unable to find any information on the UV emission from the literature. Because the exposure is shorter than that of the other sources, we extracted the counts from a $4.2'$ radius region that also includes the southern hotspot, to collect up to 99\% of the source photons. 

   \begin{figure}[h!]
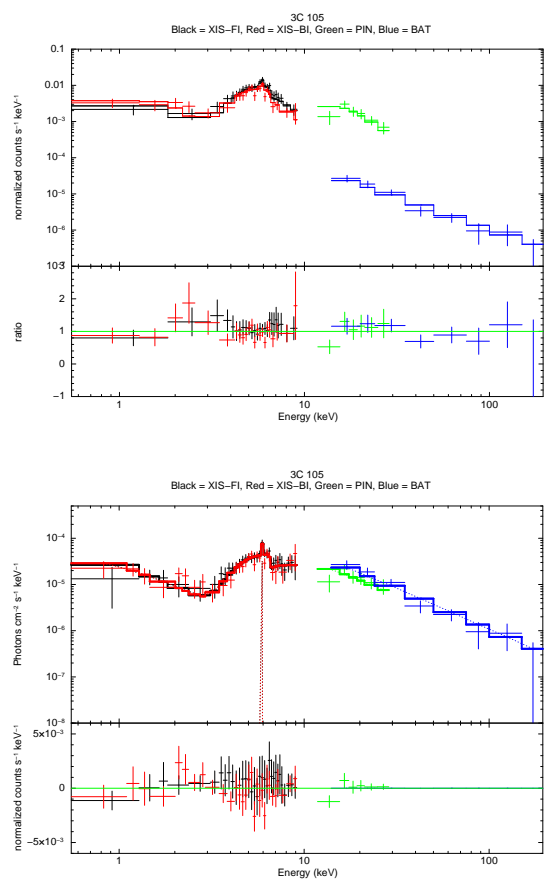

   \centering
   \includegraphics[width=.3\textwidth, angle=-90]{pap_3c105_xis_pin_allbat_spectrum.ps}\\
   \vspace{0.5cm}   
   \includegraphics[width=.3\textwidth, angle=-90]{pap_3c105_pcmA_allbat_unf_RES.eps}
   \caption{Folded (top panel) and unfolded (bottom panel) spectra and best-fit models of 3C 105: black = XIS-FI, red = XIS-BI, green = HXD/PIN, blue = Swift/BAT.}
              \label{fig:3c105_fig}%
    \end{figure}

\subsection{Spectral fitting}\label{3c105_spec}
Following the procedure reported in Sec. \ref{spectral_analysis}, the X-ray spectrum of 3C 105, given by the sum of the XIS, PIN, and BAT data, initially fitted by a simple absorbed power-law model, which leaves a strong residual in the soft energy band. The addition of thermal emission (with the \texttt{\textbf{apec}} model) requires a gas temperature that is too high ($> 2$ keV) to fit the data, while a typical value of 0.65 keV is observed in Seyfert 2 galaxies (\citealt{gua05}). The best fit is given by the partial-covering model, with the scattered power-law modeling the excess at softer energies. Using a double power-law, with different column density and photon index, the same set of spectral parameters was found. The best-fit model is \texttt{\textbf{zpcfabs*zpowerlaw + zgauss}} (Fig. \ref{fig:3c105_fig}). While the addition of Compton reflection does not improve the fit, we can evaluate an upper limit to the relative reflection. We fit the spectrum with a partially-covered \texttt{\textbf{pexrav}} model of the form \texttt{\textbf{zpcfabs*pexrav + zgauss}}. The composition and line of sight angle are settled following the model default values (solar abundance composition and 63$^{\circ}$ inclination angle). If R is equal to 0, the \texttt{\textbf{pexrav}} model translates into a partially covered power law, which is our best-fit model. Fixing the column density, scattering fraction and normalization to the best fit values, we calculated the relative reflection confidence range as a function of the power law slope (see Table \ref{tab:fit_res}, Col. 7).

\subsection{Fitting results}\label{3c105_fit_res}
Adding the nine-month-averaged BAT spectrum to the Suzaku observation, a normalization factor of 1.7 for the BAT spectrum is needed, i.e., the BAT flux is about 60\% higher than the Suzaku detection in the same band, which implies a possible long-term variability (see Sec. \ref{sec:var}).
The source 3C 105 intrinsic power-law emission is highly absorbed by an N$_{\rm H}\sim5\times10^{23}$ cm$^{-2}$, while 1.4\% of the nuclear X-ray photons is scattered at lower energies and reaches the observer unabsorbed.
\\
As previously noted, the XIS extracted region also covers the radio galaxy southern hotspot.
According to M10, its X-ray emission is much more fainter than the core: its flux is only about 30\% of the core in the 1--2 keV energy range and lower than 2\% in the 2--7 keV band. However, because we interpret the soft X-ray excess as scattered nuclear emission, we checked for a possible external contamination by comparing their fluxes. Table \ref{tab:lum} reports a summary for all sources of the flux and luminosity emitted in the main bands with (left value) and without (right value) absorption (galactic and local). In the 0.5--2 keV energy range we obtain an absorbed flux of $50\times10^{-15}$ erg cm$^{-2}$ s$^{-1}$. From M10, the total flux of the southern hotspot region in the 0.5--2 keV energy range is $5\times10^{-15}$ erg cm$^{-2}$ s$^{-1}$, ten times lower than the Suzaku/XIS observation. If there is contamination, it falls within the uncertainty of the scattering fraction. 
\\
The spectral shape of 3C 105 differs from the XIS analysis of \cite{2009ApJ...690.1322W} (W09a hereafter), especially in the power law slope and scattering fraction. This is due to the addition of the PIN data, which allows us to constrain the photon index to a steeper value. Although the source is not dominated by Compton reflection in the hard band, we obtained a reflection relative strength upper limit of 1.8, translating into a covering solid angle less than 3.6$\pi$. A narrow 136 eV EW Fe K$\alpha$ line centered at 6.4 keV is found.
\begin{table*}
\caption{ \label{tab:lum}  }
\begin{minipage}{\textwidth}
\centering
\renewcommand{\footnoterule}{}  
\begin{tabular}{c|cccc|cccc}
\multicolumn{9}{c}{\textsc{Flux and Luminosity\footnote{Fluxes and luminosity are reported with both the galactic and local absorption (left value of each column) and as intrinsic, not absorbed, emission (right value of each column) for the following energy bands: 0.5--2 keV, 2--10 keV, 10--40 keV and 14--195 keV.}}}\\
\hline
\hline
& \multicolumn{4}{c|}{\multirow{2}{*}{Flux in $10^{-12}$ erg cm$^{-2}$ s$^{-1}$}}&  \multicolumn{4}{c}{\multirow{2}{*}{Log Luminosity in erg s$^{-1}$}}\\
 &  &  &  &  &  &  &  &\\
\hline
\multirow{2}{*}{Source} & \multirow{2}{*}{0.5--2 keV} & \multirow{2}{*}{2--10 keV} & \multirow{2}{*}{10--40 keV}  & \multirow{2}{*}{14--195 keV} & \multirow{2}{*}{0.5--2 keV} & \multirow{2}{*}{2--10 keV} & \multirow{2}{*}{10--40 keV}  & \multirow{2}{*}{14--195 keV} \\
 &  &  &  &  &  &  &  &\\
\hline
\multirow{2}{*}{3C 452} & \multirow{2}{*}{0.01, 1.79} & \multirow{2}{*}{1.85, 6.59} & \multirow{2}{*}{14.46, 14.97} & \multirow{2}{*}{37.78, 38.01} & \multirow{2}{*}{41.20, 43.45} & \multirow{2}{*}{43.47, 43.88} & \multirow{2}{*}{44.36, 44.38} & \multirow{2}{*}{44.78, 44.78}\\
 &  &  &  &  &  &  &  &\\
\multirow{2}{*}{3C 105} & \multirow{2}{*}{0.05, 5.11} & \multirow{2}{*}{2.11, 8.28} & \multirow{2}{*}{9.00, 9.82} & \multirow{2}{*}{22.92, 23.26} & \multirow{2}{*}{41.99, 44.00} & \multirow{2}{*}{43.62, 44.21} & \multirow{2}{*}{44.25, 44.29} & \multirow{2}{*}{44.66, 44.66} \\
 &  &  &  &  &  &  &  &\\
\multirow{2}{*}{J0318.7+6828} & \multirow{2}{*}{0.06, 2.55} & \multirow{2}{*}{4.47, 6.03} & \multirow{2}{*}{10.23, 10.34} & \multirow{2}{*}{32.19, 32.24} & \multirow{2}{*}{42.08, 43.71} & \multirow{2}{*}{43.96, 44.09} & \multirow{2}{*}{44.32, 44.32} & \multirow{2}{*}{44.81, 44.81} \\
 &  &  &  &  &  &  &  &\\
\multirow{2}{*}{J0918.5+0425} & \multirow{2}{*}{0.01, 1.45} & \multirow{2}{*}{1.55, 2.87} & \multirow{2}{*}{4.05, 4.15} & \multirow{2}{*}{11.33, 11.37} & \multirow{2}{*}{41.84, 44.00} & \multirow{2}{*}{44.03, 44.29} & \multirow{2}{*}{44.44, 44.45} & \multirow{2}{*}{44.89, 44.89} \\ 
 &  &  &  &  &  &  &  &\\
\hline
\end{tabular}
\end{minipage}
\end{table*}

\section{Swift J0318.7+6828}
The source J0318.7+6828 is identified as an FRII GRG (see \cite{scho98} for a full description of the radio and optical properties). The optical spectrum (INT telescope in La Palma) is typical of a narrow line AGN, while the optical image shows a faint galaxy-like object associated to the radio core but the resolution is not enough to clarify the host morphology. We could not find any information on the IR or UV maps from literature, even if the IR emission, mostly in the K$_{\rm s}$ band, is found within $30''$ from the source position in the Two Micron All Sky Survey (2MASS) at IPAC. XMM--Newton observed J0318.7+6828 on February 2007 (a detailed analysis can be found in \cite{2008ApJ...674..686W}, hereafter W08): the source image, from the XMM/pn event file, in the 0.5--2 keV energy range does not show the presence of external sources and/or contamination by thermal emission. 
With a radio galaxy angular size of $15'$ (2 Mpc at z = 0.09), the giant radio lobes are well outside the Suzaku/XIS extracting region.

   \begin{figure}[h!]
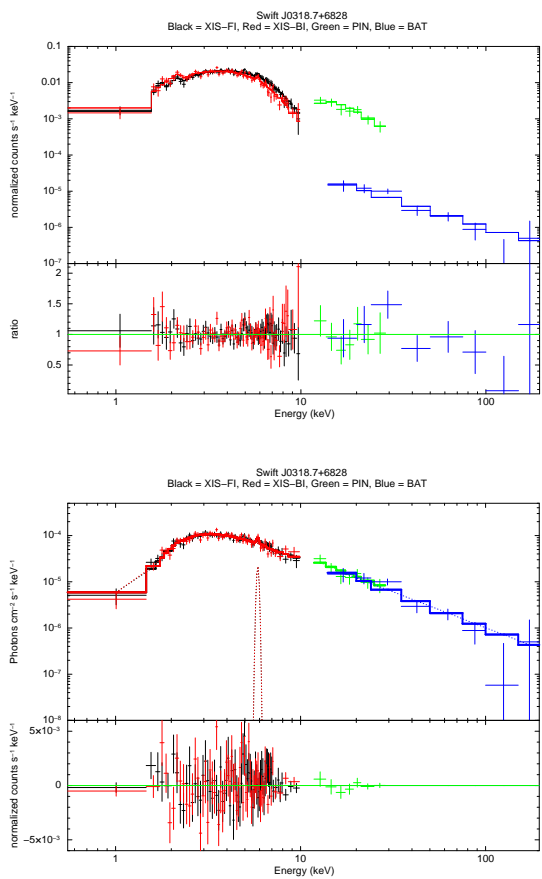

   \centering
   \includegraphics[width=.3\textwidth, angle=-90]{pap_J0318_xis_pin_allbat_spectrum.ps}\\
   \vspace{0.5cm}   
   \includegraphics[width=.3\textwidth, angle=-90]{pap_J0318_xis_pin_allbat_unf_RES.ps}
   \caption{Folded (top panel) and unfolded (bottom panel) spectra and best-fit models of Swift J0318.7+6828: black = XIS-FI, red = XIS-BI, green = HXD/PIN, blue = Swift/BAT.}
              \label{fig:J0318_fig}%
    \end{figure}

\subsection{Spectral fitting}
A simple, absorbed power law can fit the source J0318.7+6828 with sufficient accuracy, even in the soft 0.5--2 keV energy band, but the best-fit was gained by modeling its emission with a partially covered power law. A Gaussian was then added to model the faint Fe fluorescence line, fixing its rest energy at 6.4 keV, and the best-fit model takes the form \texttt{\textbf{zpcfabs*zpowerlaw + zgauss}}  (Fig. \ref{fig:J0318_fig}). No significant reflection is found in the hard part of the spectrum: we evaluated the upper limit to a possible Compton hump following the same procedure applied to 3C 105 (see Sec. \ref{3c105_spec}). 

\subsection{Fitting results}
The nuclear emission is absorbed by a column density of N$_{\rm H}\sim5\times10^{22}$ cm$^{-2}$, about an order of magnitude lower than the previous 3C sources, and the partial covering model returns a scattering factor of 1.2\%. The power-law photon index appears to be flatter than the mean value associated to Seyfert galaxies as 3C452. If reflection is present, the spectral fitting reports an upper value of 1.4 to its relative strength, i.e., the absorbing matter subtends a solid angle lower than 2.8$\pi$.

\section{Swift J0918.5+0425}\label{sec:0918}
The source Swift J0918.5+0425, the farthest among our sample (z = 0.15644) and the only radio-quiet AGN, can be defined as peculiar for many reasons. Detected by Rosat/PSPC in March 2000, the source has been identified as AGN only after the nine-month Swift/BAT survey. After being classified as a QSO Type II (\citealt{tue08}) AGN, J0918.5+0425 has been removed from the second BAT survey release because of variability, and it is not part of the 58-month catalog. The Suzaku and Swift X-ray spectra are the only X-ray data available, at present, while no additional studies in the other bands are published, except for the catalog of Swift/BAT AGN optical spectra (\citealt{2010ApJ...710..503W}, see Sec. \ref{sec:opt}).  
Its optical counterpart from the SDSS\footnote{ Sloan Digital Sky Survey (SDSS), www.sdss.org} shows a peculiar double-lobed galaxy-like object, possibly the site of a galaxy merging, but the resolution is not high enough to separate its components. 
 
   \begin{figure}[h!]
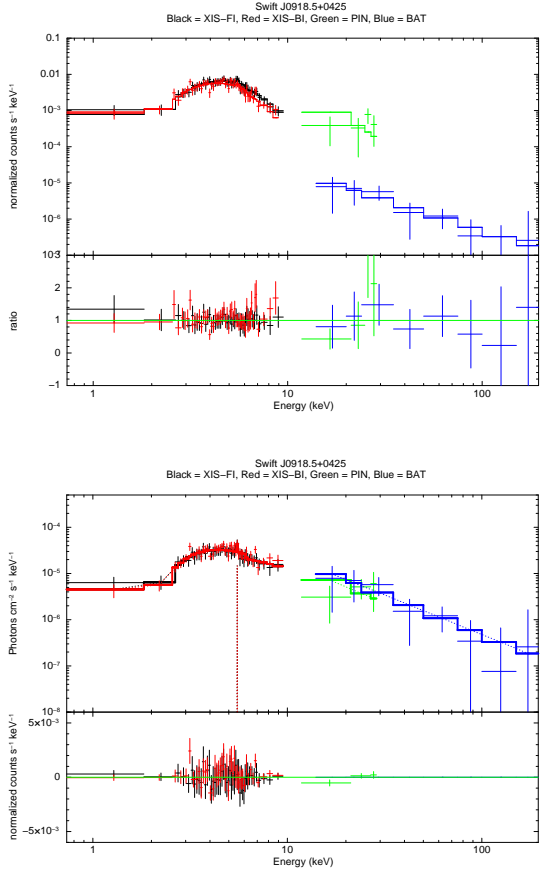

   \centering
   \includegraphics[width=.3\textwidth, angle=-90]{pap_J0918_xis_pin_allbat_spectrum.ps}\\
   \vspace{0.5cm}   
   \includegraphics[width=.3\textwidth, angle=-90]{pap_J0918_xis_pin_allbat_unf_RES.ps}
   \caption{Folded (top panel) and unfolded (bottom panel) spectra and best-fit models of Swift J0918.5+0425: black = XIS-FI, red = XIS-BI, green = HXD/PIN, blue = Swift/BAT.}
              \label{fig:J0918_fig}%
    \end{figure}

\subsection{Spectral fitting}
The PIN extracted spectrum results are not consistent with either the XIS or BAT spectra. This could be related to a low statistical significance due to the low count rate or to an incorrect particle background model, which contributes about 95\% of the total count rate. To investigate the latter hypothesis, we computed the ratio between the total detected count rate and the NXB: because the source emission is strongly affected by the particle background, we should obtain a constant value.
   \begin{figure}[h!]
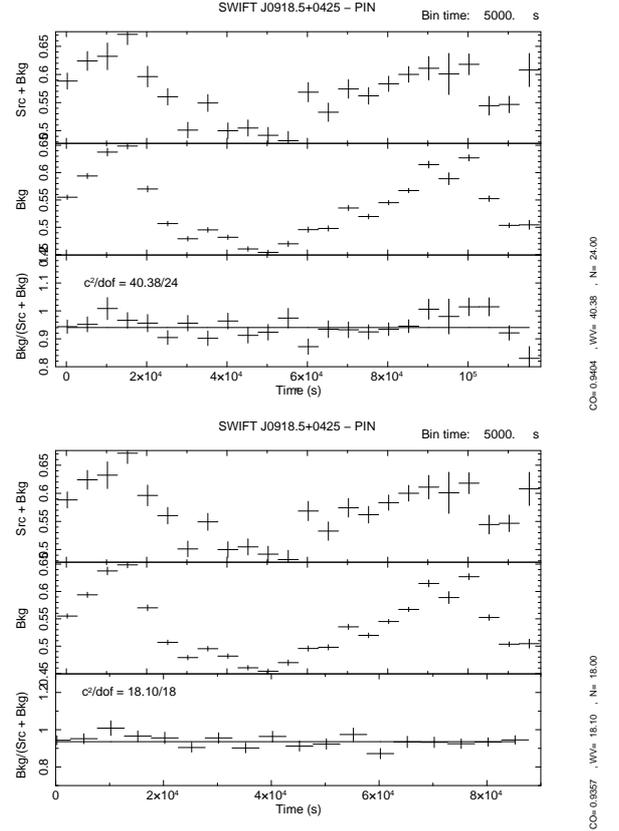

   \centering
   \includegraphics[width=.3\textwidth, angle=-90]{J0918_pin_bkg_ratio_nosel.ps}\\
   \includegraphics[width=.3\textwidth, angle=-90]{pap_J0918_pin_src_bkg_ratio_sel_lc.ps}
   \caption{Background-unsubtracted PIN (at the top), background (in the middle) light curves and their ratio ( at the bottom) of J0918.5+0425 are plotted before (top panel) and after (bottom panel) removing of the background-corrupted time interval. The goodness of the ratio fit by a constant function is also reported for the two cases.}
              \label{fig:J0918_bkg}%
    \end{figure}
\\
As shown in Figure \ref{fig:J0918_bkg} (top panel), after $9\times10^{4}$ s (spacecraft time), the ratio deviates from the constant behavior. When we remove the corrupted time interval, the PIN exposure time decreases from 52 to 43 ks, with an average count rate of 0.01 cts s$^{-1}$. The ratio resulting from the cleaned PIN light curve (Figure \ref{fig:J0918_bkg}, bottom panel) can now be well fitted ($\chi^{2}$/dof = 18.10/18) by a constant function, and the PIN spectrum is consistent with the XIS and BAT data (Figure \ref{fig:J0918_fig}). Following the same procedure as described for the other sources, the best-fit model is given by a partially covered power-law of the form \texttt{\textbf{zpcfabs*zpowerlaw}}. Adding a Gaussian profile to model the Fe K$\alpha$ emission, we can evaluate an upper limit to the line intensity, fixing the centroid rest frame energy at 6.4 keV. 

\subsection{Fitting results}
The X-ray spectrum of J0918.5+0425 was modeled by a partially covered power law, resulting in a 0.9\% scattered factor. The source is moderately absorbed (N$_{\rm H}\sim2\times10^{23}$ cm$^{-2}$ local column density) and shows a weak Fe fluorescence line (EW $< 90$ eV). We stress that this is the highest luminosity source in our sample with an intrinsic, unabsorbed 2--10 keV luminosity of $2\times10^{44}$ erg s$^{-1}$, confirming the QSO nature of the source as reported by the Swift/BAT AGN catalog.

\section{Variability}\label{sec:var}
Strong variability is found in the X-ray emission of AGN, from intrinsic changes in the source flux to variability in the spectral properties (column density, photon index) and absorbed/emitted features (\citealt{mus93}). The study of these variances and their time scales can provide information on the size and geometry of the emitting X-ray region (\citealt{mc06}), the physical conditions of the X-ray reprocessing media (\citealt{min07}) and/or the physical properties of the absorbing material close to the black hole (\citealt{mil08}).
\\
In contrast, if the line of sight to the source is blocked by obscuring matter, the direct emission is absorbed and only the photons scattered by distant, cold matter can reach the observer, losing any trace of variability. A Compton-thick AGN, where little to no high direct emission escapes below 10 keV while the hard energy part of the spectrum is dominated by reflection, can be traced by the lack of variability, both in flux and spectral shape. This is the case of Circinus, with no significant change during nine years of observations (\citealt{2009ApJ...701.1644W} and references therein), and the mildly Compton-thick source NGC 4945 (\citealt{ito08}).
\\
There are two types of variability that we can test on the basis of the Suzaku observation: variability on the time scales of the observation (ks) and variability between different observations (time scales of months/years). In the first case, we examined the light curves of both PIN and XIS detections to look for a significant variation in the source count rate, while long-term variability was studied by comparing the spectral parameters and fluxes reported here with other observations (see \cite{sob09} for a review).
Except for J0918.5+0425, we also checked for a possible variability in the BAT 14--195 keV count rate along the 66-month survey (\citealt{2011AAS...21832811B}): the discovery of variability in the energy range not dominated by reflection, i.e., the hard band observed by the BAT, in contrast with the lack of flux changes in the tens of keV, would be an additional evidence for the presence of Compton-thick material.

\subsection{Short-term variability}
\begin{figure*}[h!]
   \centering
   \includegraphics[width=.25\textwidth, angle=-90]{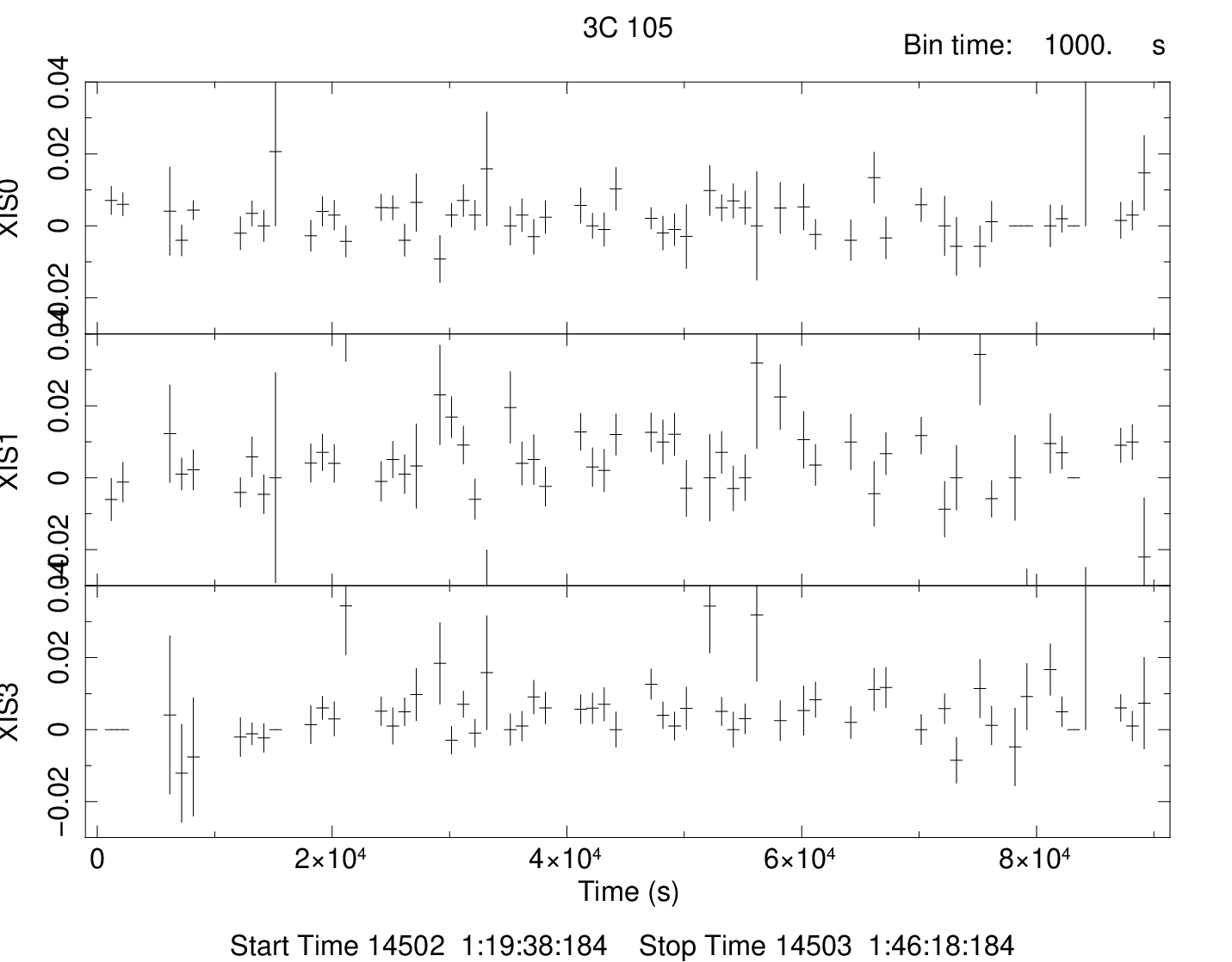}
   \includegraphics[width=.25\textwidth, angle=-90]{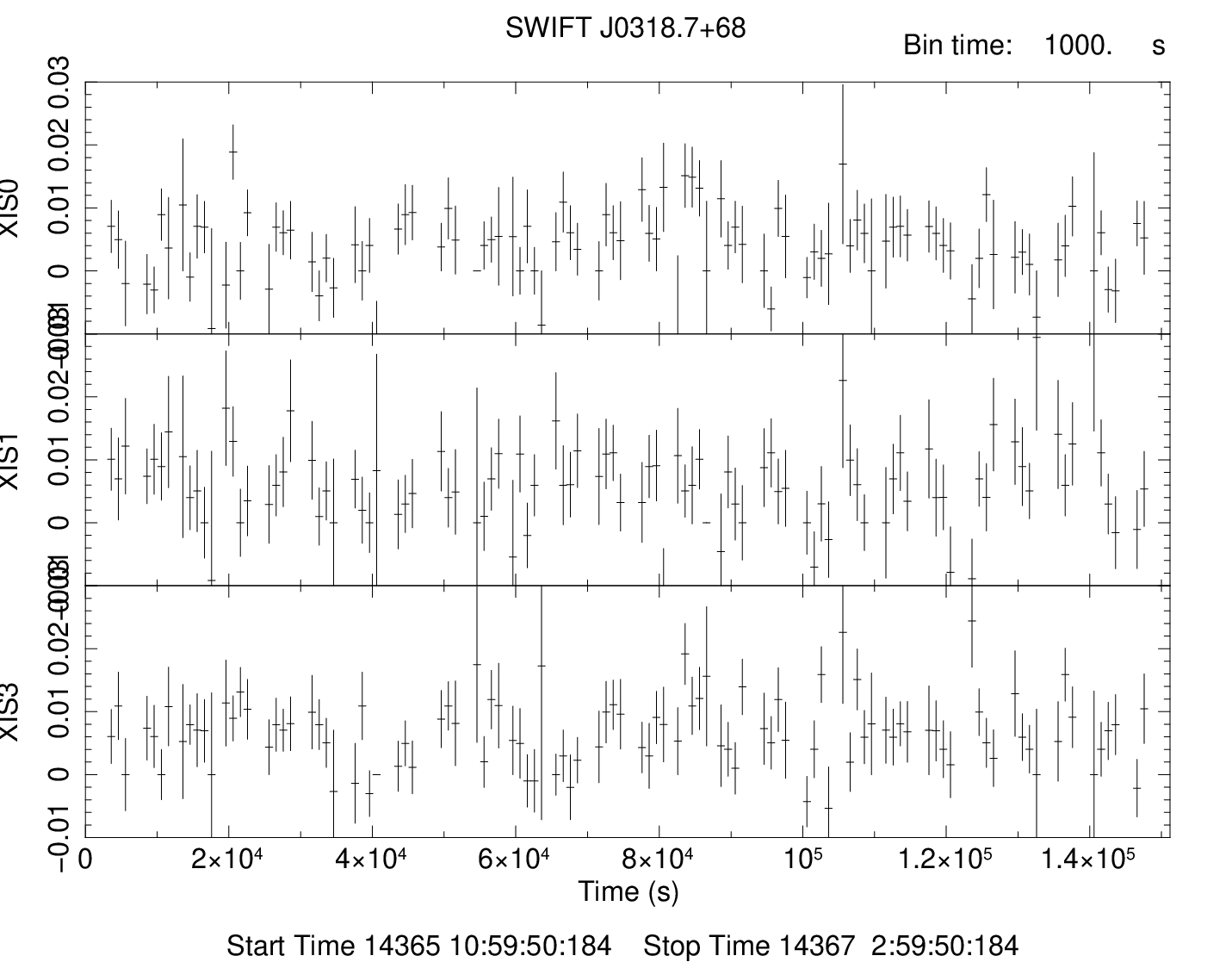}
   \includegraphics[width=.25\textwidth, angle=-90]{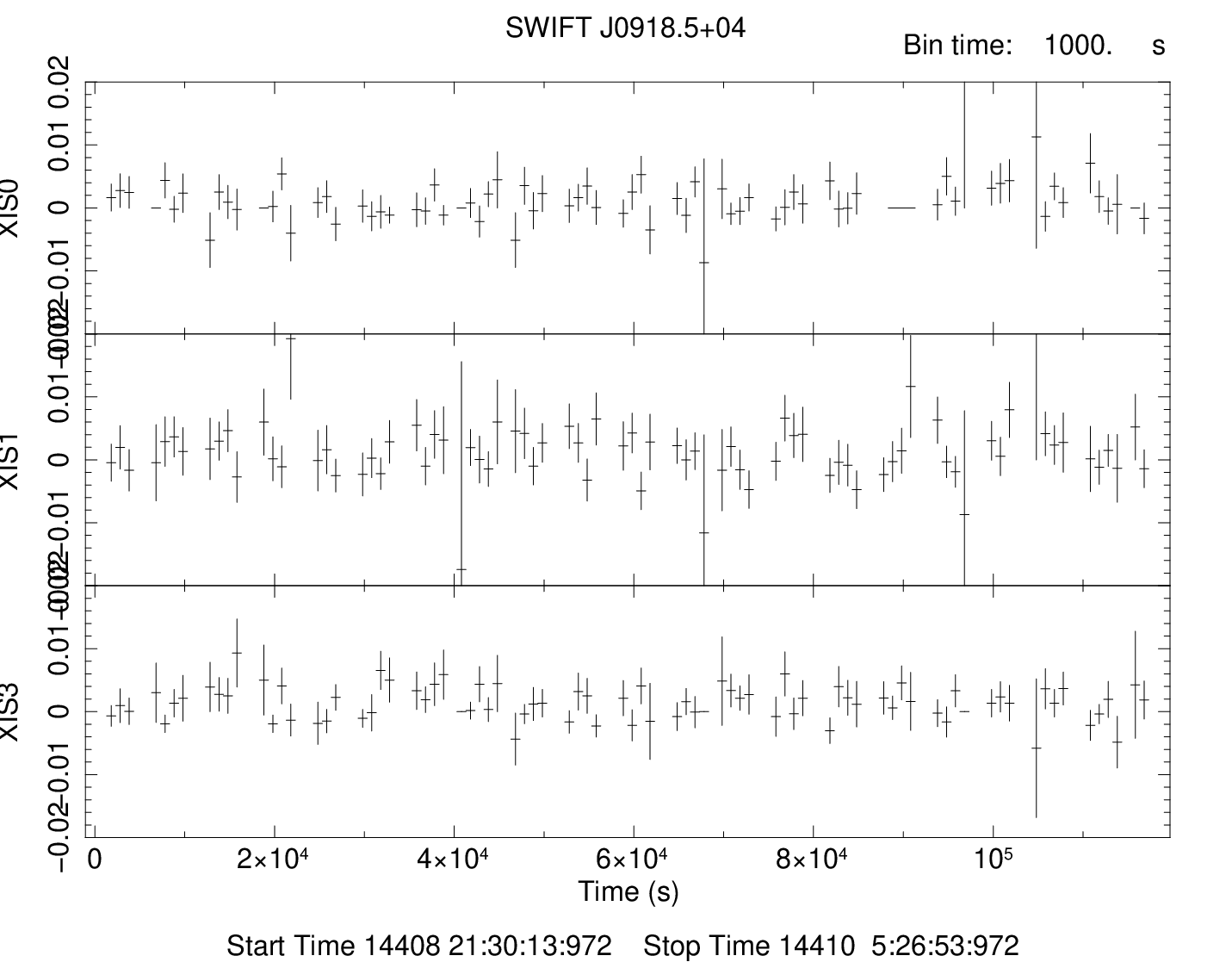}\\
   \includegraphics[width=.25\textwidth, angle=-90]{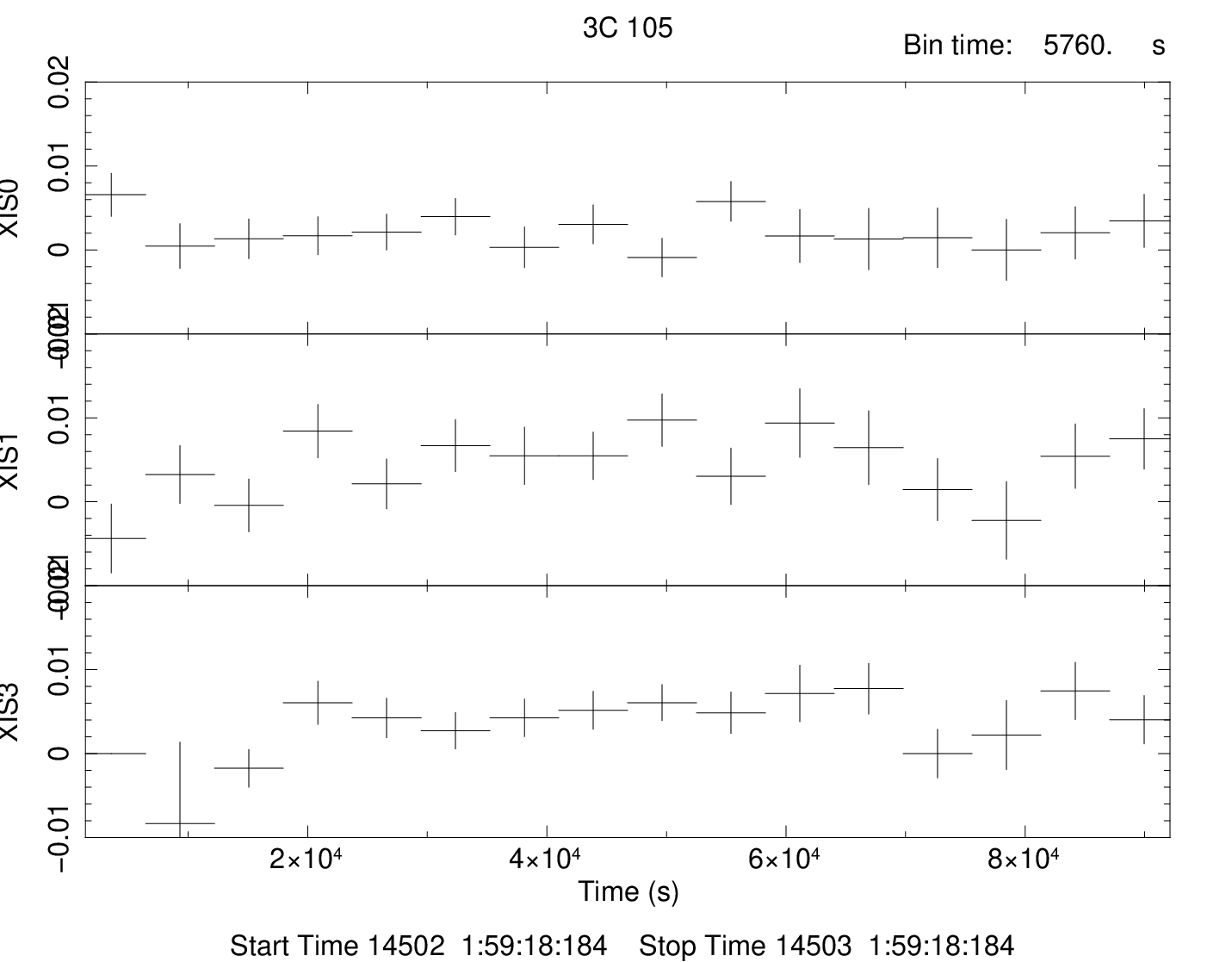}
   \includegraphics[width=.25\textwidth, angle=-90]{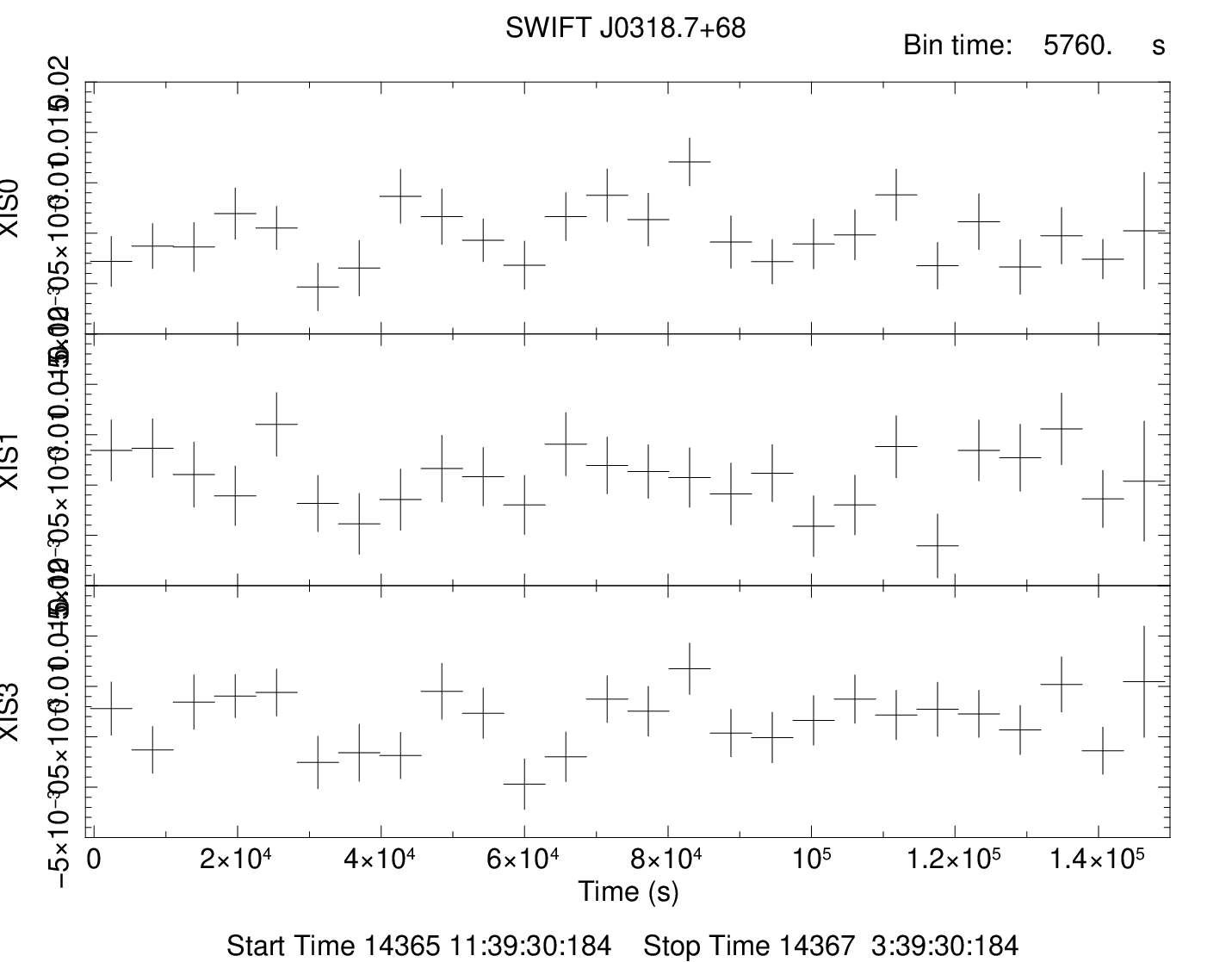}
   \includegraphics[width=.25\textwidth, angle=-90]{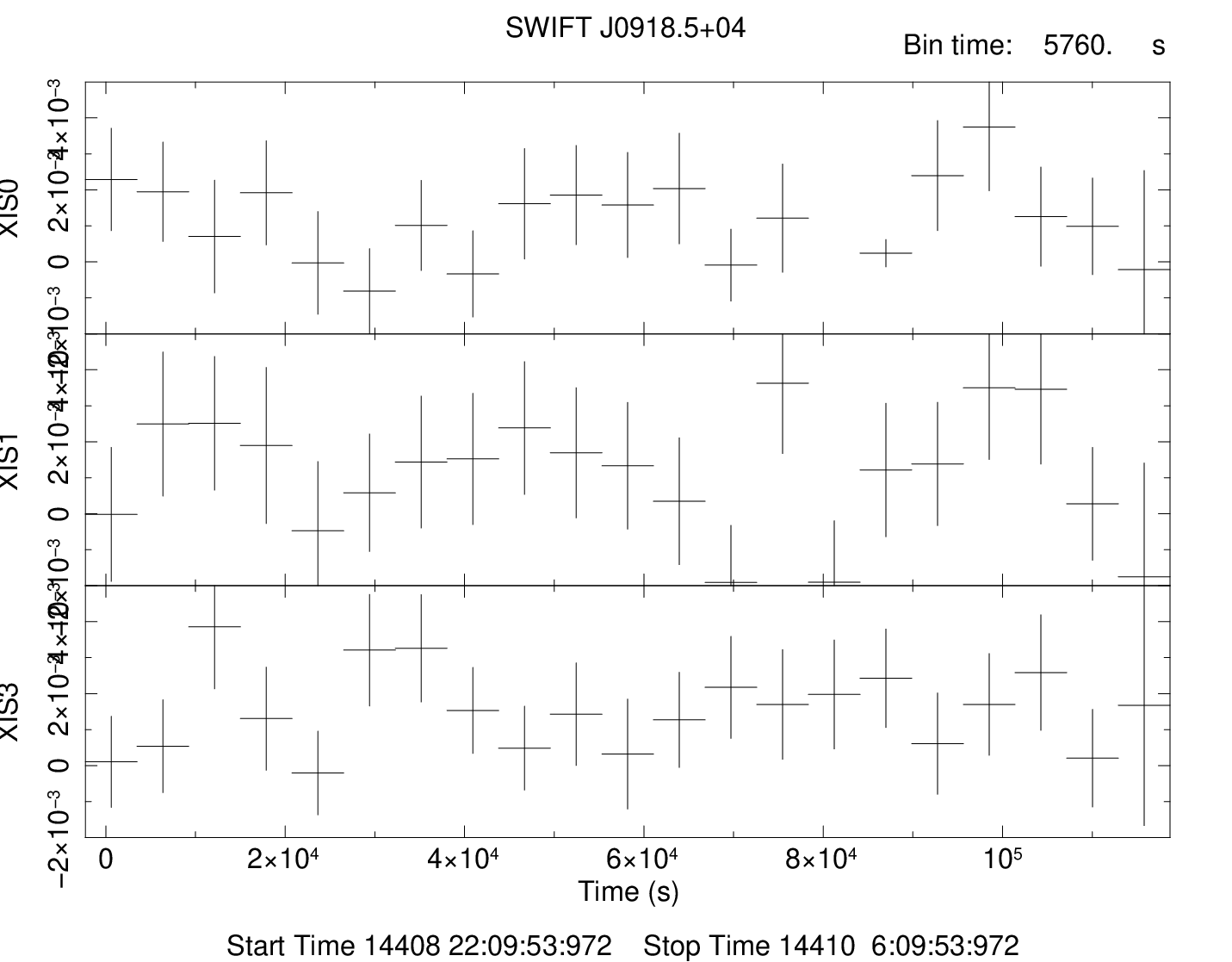}
   \caption{Suzaku/XIS light curves in the 0.5--2 keV energy range obtained with a 1 ks (top panels) and 5.76 ks (the orbital time, bottom panels) binning. The source 3C 452 is not shown given the thermal/lobes contamination in the soft band.}
              \label{fig:xis_lc_bel2}%
\end{figure*}

\begin{figure*}[h!]
   \centering
   \includegraphics[width=.19\textwidth, angle=-90]{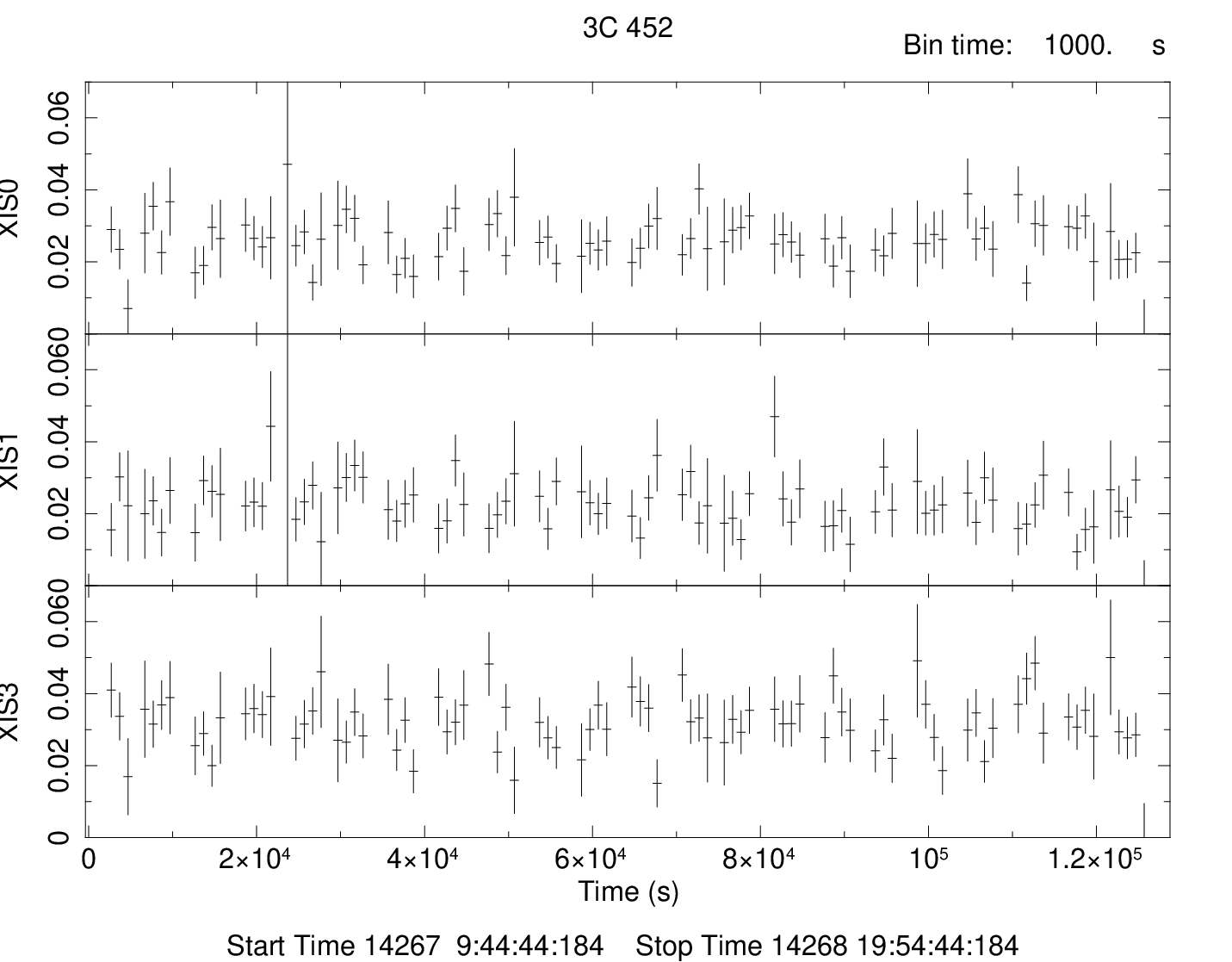}
   \includegraphics[width=.19\textwidth, angle=-90]{3c105_xis_src_and_bkg_bel2keV_1000_lc.ps}
   \includegraphics[width=.19\textwidth, angle=-90]{702075010_xis_src_and_bkg_bel2keV_1000_lc.ps}
   \includegraphics[width=.19\textwidth, angle=-90]{702076010_xis_src_and_bkg_bel2keV_1000_lc.ps}\\
   \includegraphics[width=.19\textwidth, angle=-90]{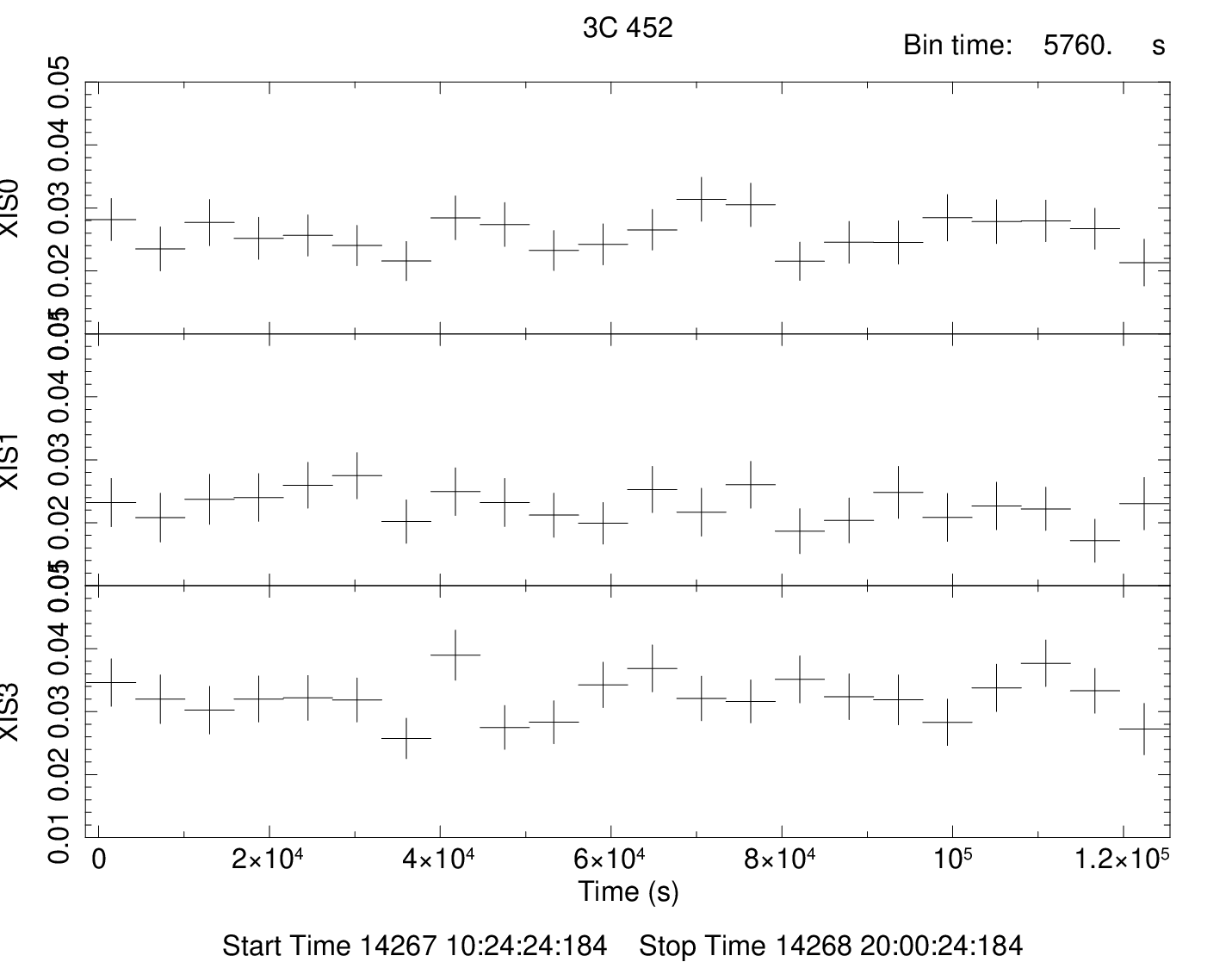}
   \includegraphics[width=.19\textwidth, angle=-90]{3c105_xis_src_and_bkg_bel2keV_lc.ps}
   \includegraphics[width=.19\textwidth, angle=-90]{702075010_xis_src_and_bkg_bel2keV_lc.ps}
   \includegraphics[width=.19\textwidth, angle=-90]{702076010_xis_src_and_bkg_bel2keV_lc.ps}
   \caption{Suzaku/XIS light curves in the 2--10 keV energy range, except for 3C 452 cutted below 3 keV because of contamination, obtained with a 1 ks (top panels) and 5.76 ks (the orbital time, bottom panels) binning.}
              \label{fig:xis_lc_abo2}%
\end{figure*}

The XIS light curves, plotted with a 1 ks and 5.76 ks (the orbital time scale) binning, were extracted in two energy ranges, 0.5--2 keV (Figure \ref{fig:xis_lc_bel2}) and 2--10 keV (Figure \ref{fig:xis_lc_abo2}). This allowed to explore the variability of the soft X-ray emission.
\\
For 3C 452, XIS light curves were only extracted in the 3--10 keV band because of contamination in the soft, extended emission. The PIN is plotted with a 5.76 ks binning along with the NXB light curve because it is strongly affected by the flaring behavior of the particle background. We do not show the PIN light curve for J0918.5+0425 since the NXB contributes the 97.6\% of the observation and a longer exposure is needed to evaluate a possible variability.
\begin{figure*}[h!]
   \centering
   \includegraphics[width=.23\textwidth, angle=-90]{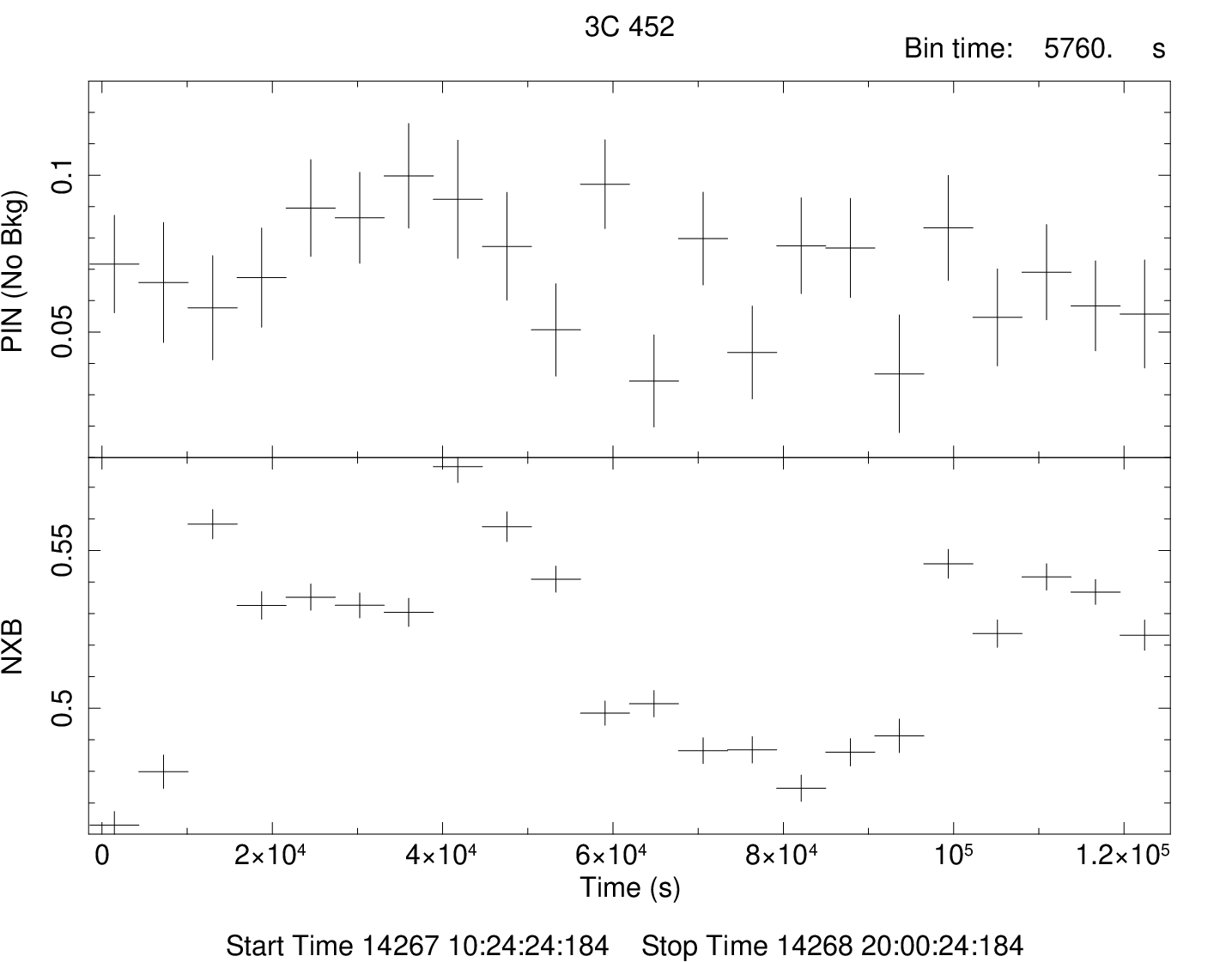}
   \includegraphics[width=.23\textwidth, angle=-90]{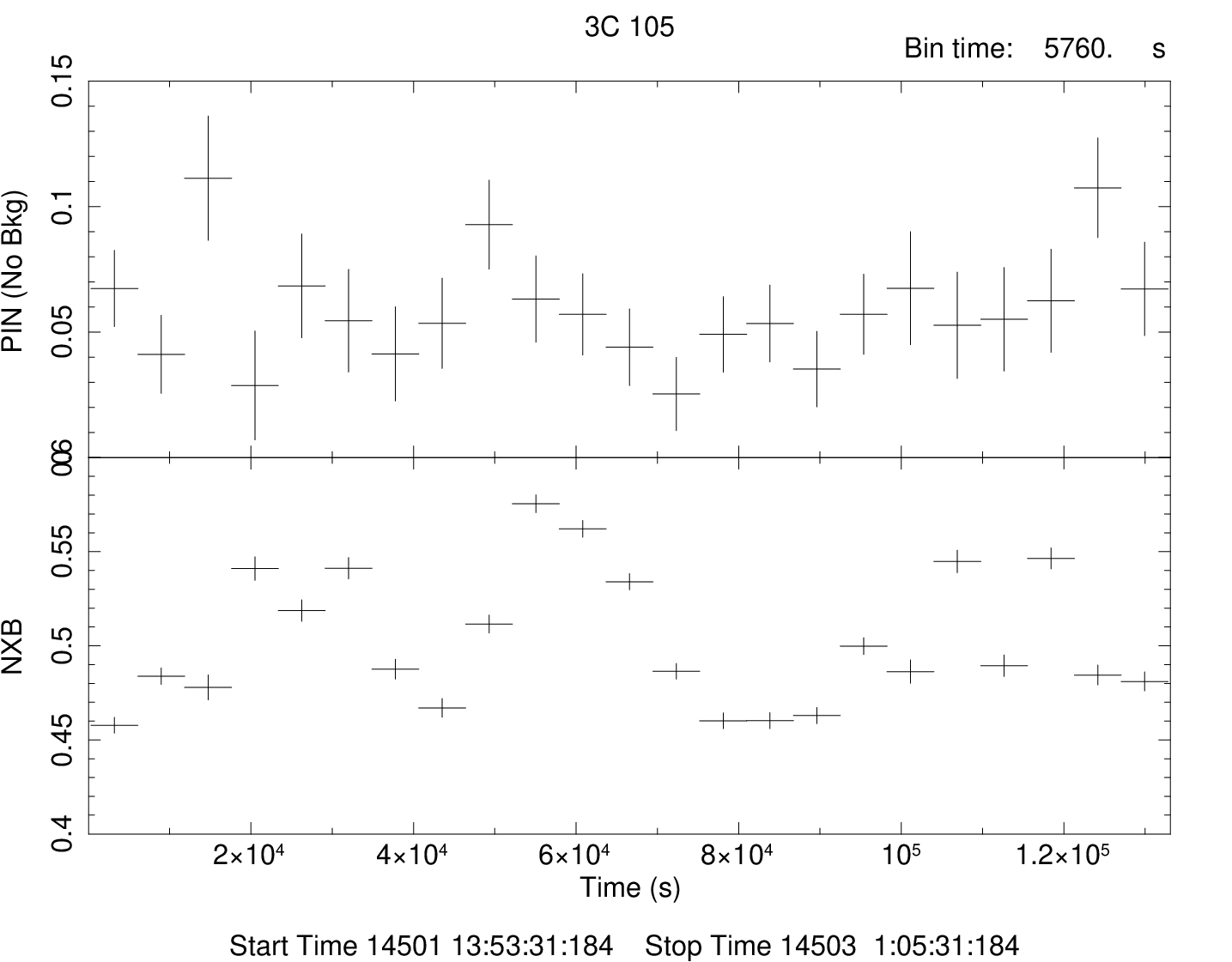}
   \includegraphics[width=.23\textwidth, angle=-90]{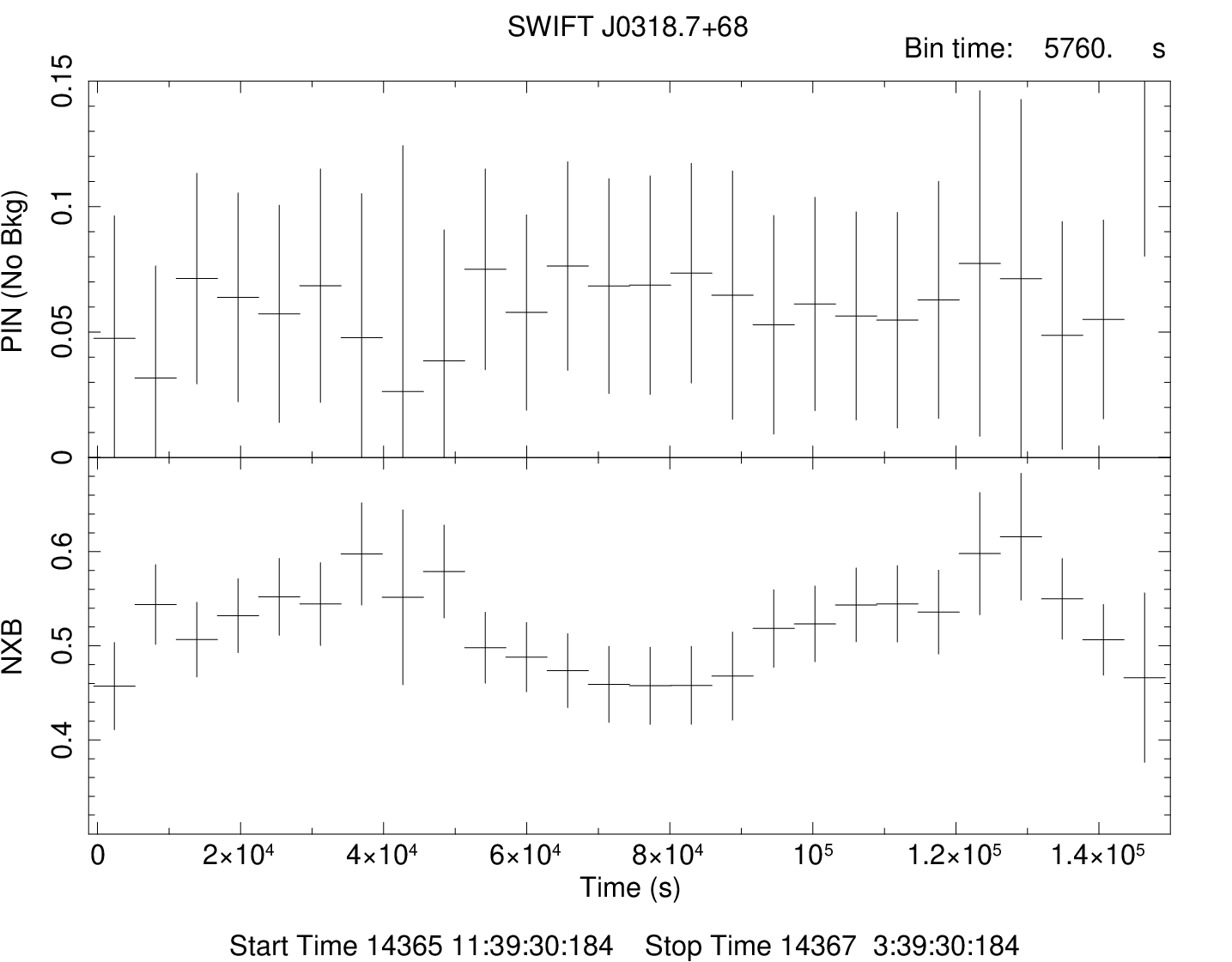}
   \caption{Background-subtracted PIN and NXB light curves above 10 keV, except for J0918.5+0425 where the 97.6\% of the detection is affected by the NXB and a longer exposure is needed to evaluate a possible intensity variability.}
              \label{fig:pin_lc}%
\end{figure*}
All XIS light curves were fitted by a constant function using the \texttt{QDP} fitting suite. No variability was found in the soft band but, given the low level of counts, this result cannot be unambiguously related to the reprocessed nature of the emission. Above 2 keV, we found a slight deviation from a constant trend for the sources 3C 105 and J0318.7+6828 in the 5.76 ks binning, where we also obtained the lowest upper limit for reflection. We confirm that no strong, significant short-term variability (factor of 2 or higher) is found below 10 keV. The PIN light curves instead show a remarkable count rate temporal variation, which, however, reproduces the NXB fluctuations: the temporal behavior of the PIN is clearly affected by the particle background, hiding any potential AGN variability.

\subsection{Long-term variability}
The spectral properties obtained from the analysis of Chandra, XMM-Newton, and Swift/XRT data were collected from the literature (see Table \ref{tab:other_obs}). For 3C 105, a simple model to which we applied, differences in the photon index or column density could fall within the uncertainty of our best-fit model. The X-ray spectral analysis for other sources has not been published yet (e.g., the XMM-Newton observation of 3C 452).
While a detailed analysis of these observations is not the aim of this work, we compared our results with those from Suzaku by fitting these spectra with our best-fit model, letting the normalization free to vary for changes in luminosity and adding the nine-month-averaged BAT data to constrain the photon index. When the source could not be described by our best-fit parameters, we looked for intrinsic variations in the spectral parameters, with the assumption that our best-fit model reflects the physical properties of the AGN. In the case of the XRT data analysis the Fe K line was not modeled because of the low-energy resolution of the instrument.
The Swift/XRT data filtering was performed by means of the HEASARC \texttt{xrtpipeline} following the standard event screening criteria using the CALDB release of the observation epoch. The Chandra/ACIS analysis is based on the cleaned event files of the Chandra archive third reprocessing (ASCDSVER = 7.6.11.3, CALDB = 3.4.2). For the XMM-Newton/pn data, we reduced the data following the standard procedures as adviced by the XMM-Newton SOC using \texttt{SAS v.10.0.0}. The extracting regions range from $50''$--$60''$ for the XRT images to $30''$ and $7''$ for XMM-Newton and Chandra images, respectively.
\begin{table*}
\caption{ \label{tab:other_obs}  }
\begin{minipage}{\textwidth}
\begin{tiny}
\centering
\renewcommand{\footnoterule}{}
\begin{tabular}{c|cccccccccc}
\multicolumn{11}{c}{\small{\textsc{Spectral Fit Results from Chandra/XMM-Newton/Swift XRT}}}\\
\hline
\hline
 &  &  &  &  &  &  &  &  &  &\\
\multirow{2}{*}{\small{Source}} & \multirow{2}{*}{\small{Instr.\footnote{Observatory name: S = Suzaku/XIS, B = Swift/BAT, C = Chandra/ACIS, X = XMM-Newton/EPIC, SX = Swift/XRT.}}} & \multirow{2}{*}{\small{Obs. Date}} & \multirow{2}{*}{\small{Model\footnote{The best-fit models: A = N$_{\rm H}$*PL, B = N$_{\rm H}$*PL + Gauss, C = Pcfabs*PL + Gauss, D = Pcfabs*PL + Gauss + Pexrav, E = N$_{\rm H}$*PL + Gauss + Pexrav + Th, F = Pcfabs*PL + Gauss + Pexrav + Th. All models include the galactic absorption, not shown here for simplicity.}}} & \small{N$_{\rm H}$\footnote{Source-intrinsic column density and primary power-law photon index (p indicates a value pegged at the lower/upper limit of 1.5/1.9).}}& \multirow{2}{*}{\small{$\Gamma$}$^{\rm c}$} & \multirow{2}{*}{\small{f$_{\rm scatt}$\footnote{Where modeled, the fraction of the scattered component relative to the intrinsic power law.}}} & \small{Fe K E, EW\footnote{Centroid energy and equivalent width with respect to the whole continuum of the Fe fluorescence K$\alpha$ line at the rest frame of the source redshift.}} &  \multirow{2}{*}{\small{Refl.\footnote{Y = present, N = not found, / = not modeled}}} & Log L$_{[2-10 \rm\; keV]}$\footnote{(obs) = observed, (unabs) = absorption corrected, (pow) = power law.} & \multirow{2}{*}{\small{Ref.\footnote{1 = \cite{iso02}, 2 = \cite{eva06}, 3 = \cite{2009ApJ...690.1322W}, 4 = here.}}} \\
&  &  &  & [$10^{22}$ cm$^{-2}$] & & & [keV] &  & [10$^{44}$ erg s$^{-1}$] & \\
 &  &  &  &  &  &  &  &  &  & \\
\hline
 &  &  &  &  &  &  &  &  &  &\\
\multirow{4}{*}{\small{3C 452}} & C & 2002-08-27 & E & ${59}^{+9}_{-9}$ &  ${1.66}^{+0.28}_{-0.21}$& / & ${6.40}^{+1.0}_{-0.7}$, $<0.22$ & Y & 0.7(unabs) & 1\\
& C & 2002-08-27 & E  & ${57}^{+9}_{-8}$ & ${1.7}$(f) & / & ${6.4}$(f), ${0.1}$(f)& Y & 1.(pow) & 2 \\
& S+B & 2007-06-16 & D  & ${43.5}^{+10.9}_{-6.9}$ & ${1.55}^{+0.14}_{-0.11}$ & $<0.5\%$ &${6.43}^{+0.03}_{-0.03}$, 0.164 & Y & 0.8(unabs) & 4 \\
& X+B & 2008-11-30 & F  & ${63.9}^{+8.5}_{-7.2}$ & ${1.5}$(p) & $<0.9\%$ & ${6.44}^{+0.03}_{-0.03}$, ${0.15}$& Y & 1.1(unabs) & 4 \\
 &  &  &  &  &  &  &  &  &  & \\
\multirow{4}{*}{\small{3C 105}} & SX+B & 2006-07-11/16 & C & ${38.6}^{+4.9}_{-4.0}$ & 1.9(p) & $<0.2\%$ & / & N & 3.4(unabs) & 4 \\
& C+B & 2007-12-17 & C & ${52.6}^{+7.3}_{-5.9}$ & 1.9(p) & $<0.2\%$ & 6.4(f), 0.09 & N & 2.7(unabs) & 4\\
& S+B & 2008-02-05 & C & ${45.9}^{+6.2}_{-6.6}$ & ${1.78}^{+0.20}_{-0.19}$ & ${1.4}^{+0.9}_{-0.5}\%$ & ${6.40}^{+0.07}_{-0.05}$, 0.136 & N & 1.6(unabs) & 4\\
& X+B & 2008-02-25 & C & ${48.4}^{+14}_{-10}$ & ${1.9}$(p) & $<0.5\%$ & ${6.4}$(f), 0.09 & N & 1.8(unabs) & 4\\
 &  &  &  &  &  &  &  &  &  & \\
\multirow{4}{*}{\small{J0318.7+6828}}  & X & 2006-01-29 & C & ${4.10}^{+0.48}_{-0.41}$ & ${1.52}^{+0.12}_{-0.11}$ & ${3.3}^{+0.9}_{-0.9}\%$ & 6.4(f), ${0.044}^{+0.04}_{-0.04}$ & N & 1.5(obs) & 3 \\
& SX & 2006-03-29 & A & ${3.70}^{+1.74}_{-1.83}$ & ${1.73}^{+0.48}_{-0.55}$ & / & / & N & 0.8(obs) & 3 \\
& SX & 2006-04-05 & A & ${3.66}^{+1.44}_{-1.41}$ & ${1.44}^{+0.40}_{-0.46}$ & / & / & N & 1.0(obs) & 3 \\
& S+B & 2007-09-22 & C & ${5.26}^{+0.42}_{-0.41}$ & ${1.72}^{+0.20}_{-0.19}$ & ${1.2}^{+0.6}_{-0.4}\%$ & 6.4(f), 0.063 & N & 1.2(unabs) & 4 \\
 &  &  &  &  &  &  &  &  &  & \\
\hline
\end{tabular}
\end{tiny}
\end{minipage}
\end{table*}%

\subsubsection{3C 452}
The spectral parameters of 3C 452 agree well with the detailed analysis of the Chandra observation performed by I02 and \cite{eva06}. They also confirm the reflection component and the narrow Fe K feature. We included the thermal emission as modeled by I02 (accounted for by the normalization of the extended emission in the Suzaku analysis) and the scattering component in the analysis of the XMM-Newton observation (the Swift/XRT detection was not available). A partially covered power law coupled with strong reflection gives a good fit ($\chi^{2}/\rm d.o.f$ = 296/291) to the data for the XIS analysis. We found a low fraction of scattered emission ($<0.9\%$) with an uncertainty due to the presence of thermal photons. The spectral properties of 3C 452, i.e., high column density, flattened photon index, narrow Fe K line, and strong reflection, do not show variability during six years. 
This results, coupled with an unchanged absorption-corrected 2--10 keV luminosity, confirm 3C 452 as a highly absorbed, reflection-dominated source.

\subsubsection{3C 105}
The X-ray analysis of 3C 105 has been published by \cite{aje08} (Swift/XRT + BAT) and \cite{mas10} (Swift/XRT, Chandra/ACIS and XMM-Newton/pn). While these authors generally confirm the strong absorption ($>3\times10^{23}$ cm$^{-2}$), the use of simple absorbed power laws to model the spectrum and the lack of information regarding the source luminosity did not allow them to correctly evaluate a possible intrinsic variability. After summing the four available Swift/XRT observations, which span a time interval of five days, we try to describe the Swift/XRT, Chandra/ACIS and XMM-Newton spectra with the best-fit model and parameters obtained from the Suzaku spectral analysis (see Sec. \ref{3c105_fit_res}). The source is intrinsically variable, and the new best-fit parameters are listed in Table \ref{tab:other_obs} for each observation. The lack of reflection and the narrow Fe K line are confirmed but the scattering fraction is lower than 0.2\% for Chandra and Swift observations, and lower than 0.5\% for the XMM-Newton detection. These changes are coupled to an absorption-corrected 2--10 keV luminosity that decreases by more than 50\% from the middle of 2006 to the first months of 2008. A slight variability ($\sim10\%-20\%$) in the column density is also found.
   \begin{figure}[h!]
   \centering
   \includegraphics[width=.37\textwidth, angle=-90]{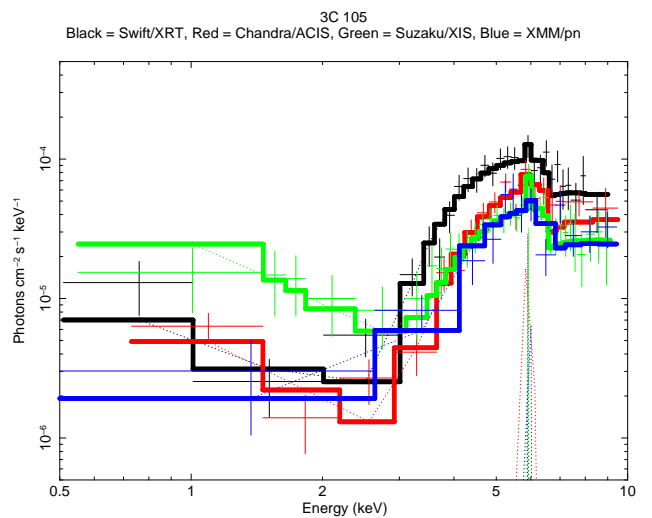}\\
   \caption{Unfolded spectrum of 3C 105 as observed by Swift/XRT (2006-07-11/16, black line), Chandra/ACIS (2007-12-17, red line), Suzaku/XIS (2008-02-05, green line), and XMM-Newton (2008-02-25, blue line).}
              \label{fig:3c105_var}%
    \end{figure}
The intrinsic variable nature of 3C 105 is clearly visible in Fig. \ref{fig:3c105_var}, where the unfolded spectra of the four observations are plotted together: a higher intrinsic luminosity is associated to lower values of the scattering fraction.

\subsubsection{Swift J0318.7+6828}
A detailed analysis and comparison of the Swift/XRT and XMM-Newton observations of J0318.7+6828 is reported in W08: flux variability is found, with the XMM detection showing the source to be brighter by $\sim30\%$ two months before the two XRT observations. Suzaku observed the source more than one and a half years later: column density, power law slope and the Fe K$\alpha$ feature are consistent with the previous results, but the scattered emission at lower energies is significantly lower than the XMM-Newton observation. The observed luminosity in the 2--10 keV resulting from our analysis is $0.9\times10^{44}$ erg s$^{-1}$, consistent with the values given by the XRT observation. We confirm the long term variability reported by W08: the source is seen to be more luminous and more scattered in early 2006, followed by a decrease in X-ray power of about 30\% after two months(no information is available for the scattering component in the XRT data), and the same luminosity is found more than one year later. The scattering factor shows a strong decrease from the XMM to the Suzaku observation.

\subsubsection{Swift J0918.5+0425}
As previously noted, the Suzaku observation of J0918.5+0425 represents the first attempt to study its broad band X-ray emission after the discovery of its high hard X-ray luminosity in the BAT survey. Except for a detection in the soft band by Rosat, the Swift/XRT view J0918.5+0425 is the only available observation one can use to look for variability. Unfortunately, the count level is low (46 counts in the whole band) and it is not statistically significant below 3 keV. We were able to fit the data with our best-fit model after removing the Gaussian profile. However, this is probably due to the uncertainty of the XRT detection and not to the lack of variability. Indeed, the flux in the 2--10 keV band is about 50\% lower than the XIS observation and almost four times lower than the nine-month averaged BAT flux extrapolated in the same energy range. We conclude that the source shows long-term variability, but longer exposures are needed for a detailed analysis.

\subsection{BAT 66-month lightcurves}
   \begin{figure}[h!]
   \centering
   \includegraphics[width=.45\textwidth]{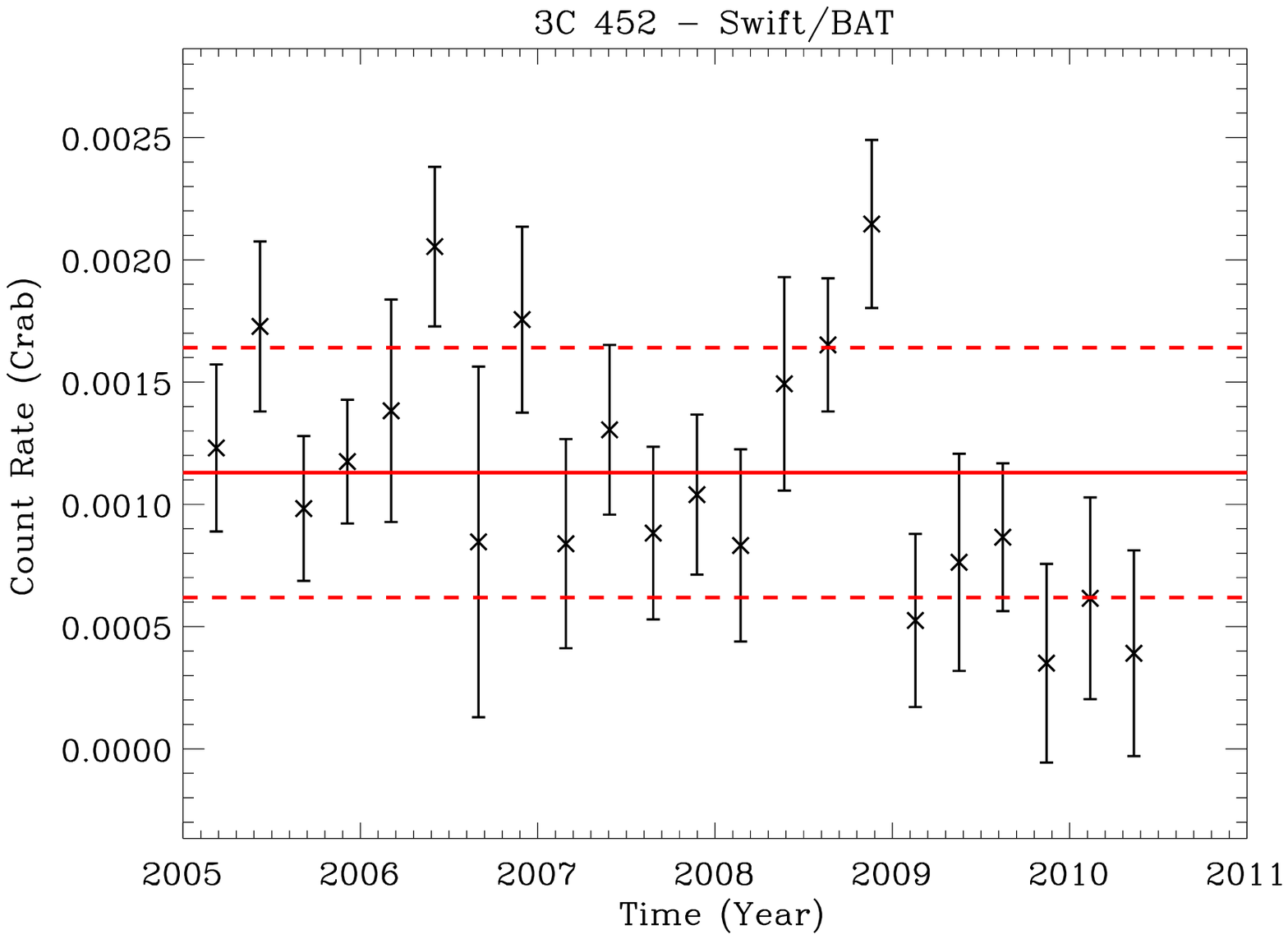}
   \vspace{0.1cm}   
   \includegraphics[width=.45\textwidth]{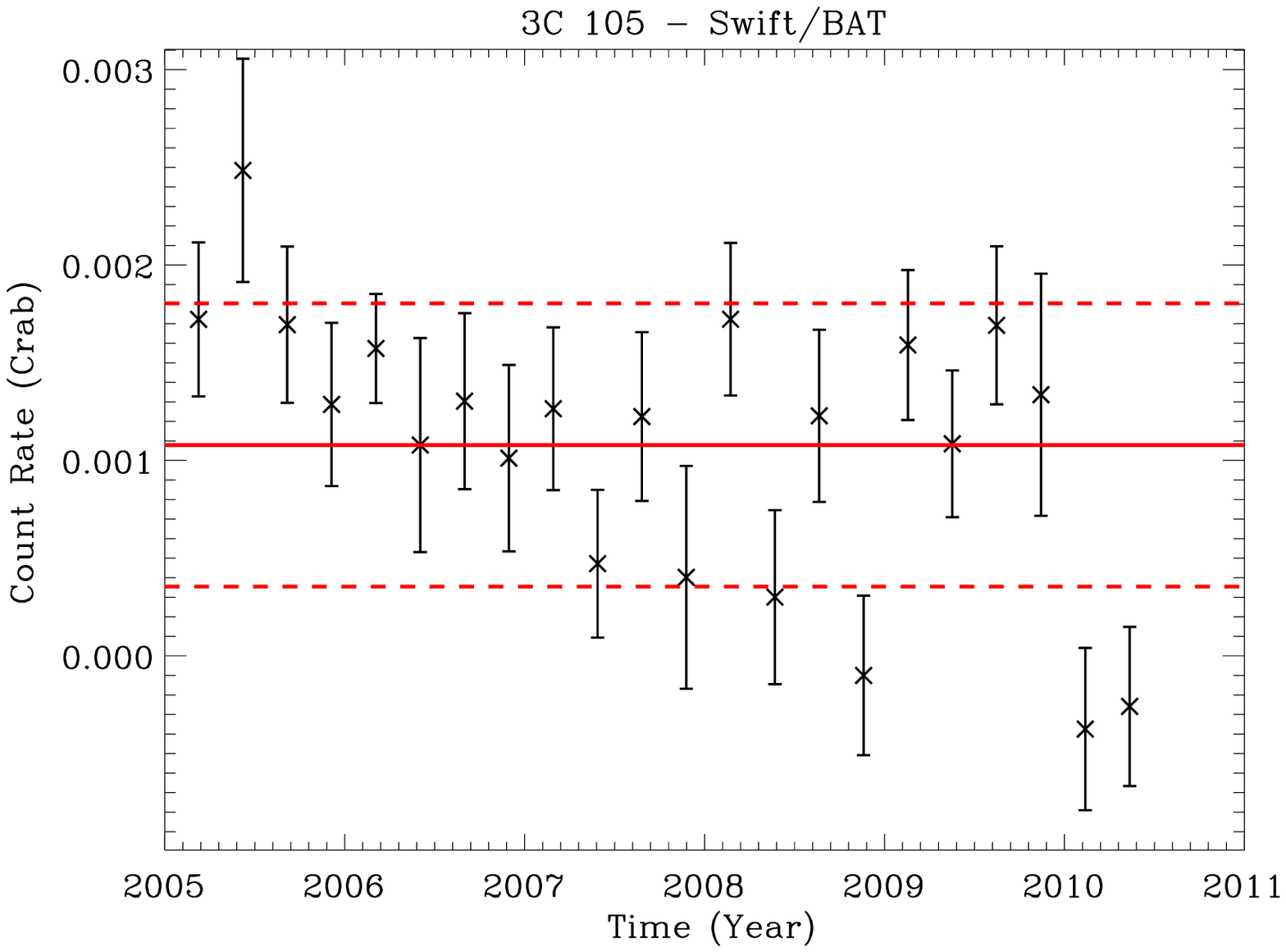}
   \vspace{0.1cm}   
   \includegraphics[width=.45\textwidth]{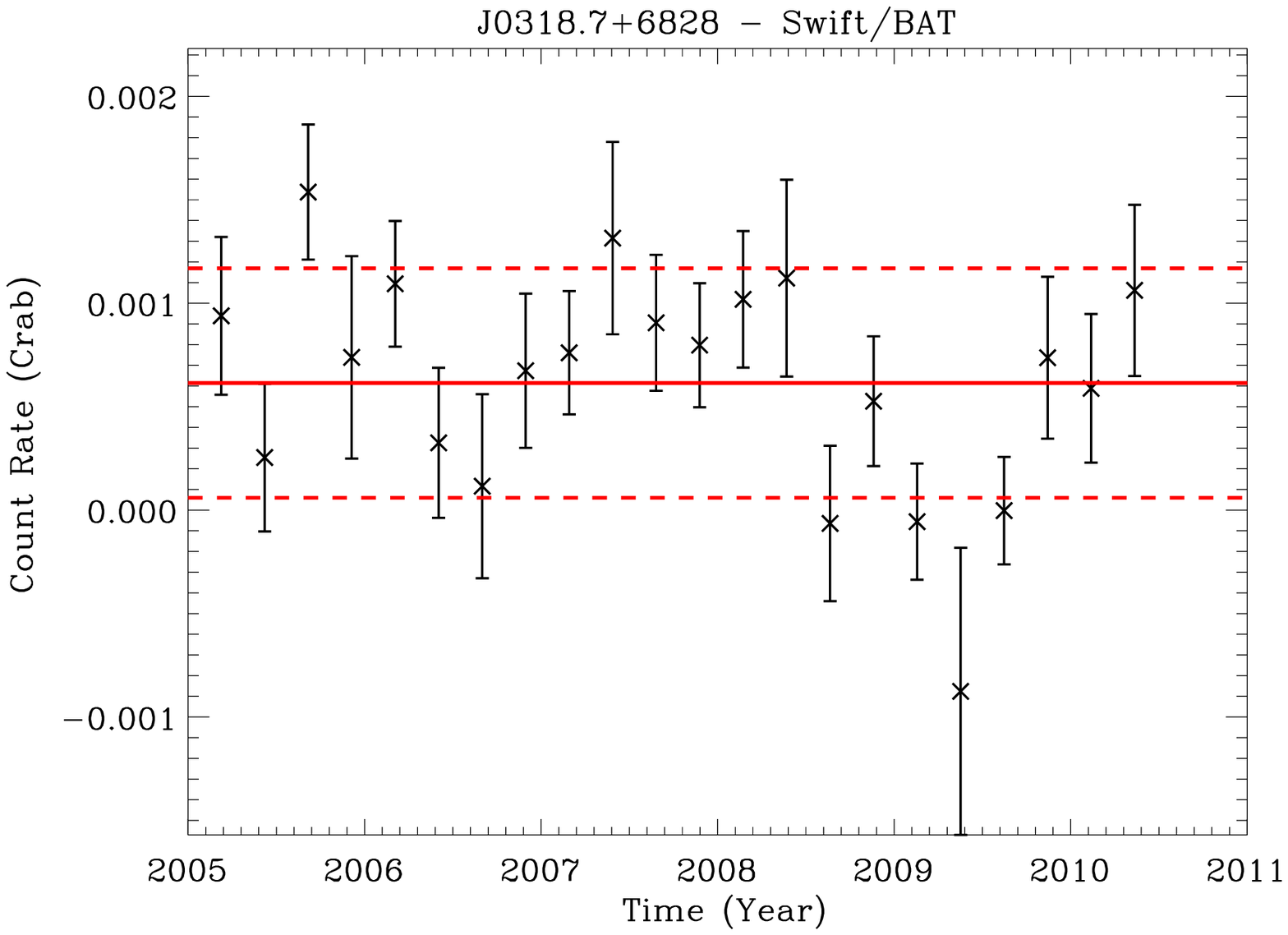}
   \caption{Swift/BAT 66-month Crab weighted light curves of 3C 452 (top panel), 3C 105 (central panel), and J0318.7+6828 (bottom panel), plotted with a three-month binning. The horizontal continuous and dashed lines refer to the average rate and the $\pm1\sigma$ standard deviation.}
              \label{fig:batlight}%
    \end{figure}
The Swift/BAT 66-month Crab weighted light curves in the 14--195 keV energy range of 3C 452, 3C 105, and J0318.7+6828 (J0918.5+0425 was removed from the BAT AGN catalog) were binned with the \texttt{rebingausslc} tool using a three-month bin, and are plotted in Figure \ref{fig:batlight}. The horizontal continuous and dashed lines refer to the average rate and the $\pm1\sigma$ standard deviation. Although the count rate of our sample, especially 3C 452, seems to increase/decrease in a coherent way, all points and the associated error bars lie within $2\sigma$.

\section{Scattering fraction}\label{sec:opt}
\begin{table*}
\caption{ \label{tab:opt}  }
\begin{minipage}{\textwidth}
\centering
\renewcommand{\footnoterule}{}
\begin{tabular}{c|ccccccccc}
\multicolumn{10}{c}{\textsc{Optical Properties of the Sample\footnote{De-reddened emission line properties taken from \cite{2010ApJ...710..503W}, based on observations from the 2.1 m Kitt Peak National Observatory (KPNO) telescope except for J0918.5+0425, which is based on archival SDSS data. All luminosities are expressed as logarithms in units of erg s$^{-1}$.}}}\\
\hline
\hline
\multirow{2}{*}{Source} & \multirow{2}{*}{Obs. Date} & \multirow{2}{*}{E(B--V)$_{\rm gal}$\footnote{Milky Way and source-intrinsic extinction.}} & \multirow{2}{*}{E(B--V)$_{\rm int}\hspace{0.3 mm}^{b}$} & \multirow{2}{*}{L(H$\alpha$)} & \multirow{2}{*}{L(H$\beta$)}& \multirow{2}{*}{L[OIII]$\lambda$5007} & \multirow{2}{*}{L[OIII]/L(H$\beta$)} & \multirow{2}{*}{L$_{[2-10 \;\rm keV]}$\footnote{Logarithm of the intrinsic 2--10 keV power law luminosity in erg s$^{-1}$.}} & \multirow{2}{*}{R$^{\rm X}_{\rm [OIII]}$\footnote{The ratio of the intrinsic 2--10 keV power law luminosity with respect to the corrected L[OIII].}} \\
 & & & & & & & & & \\
\hline
\multirow{2}{*}{3C 452} & \multirow{2}{*}{2006-11-21} & \multirow{2}{*}{0.14} & \multirow{2}{*}{0.84} & \multirow{2}{*}{40.94} & \multirow{2}{*}{40.09} & \multirow{2}{*}{40.89} & \multirow{2}{*}{6.31} & \multirow{2}{*}{43.81} & \multirow{2}{*}{832} \\
 & & & & & & & & & \\
\multirow{2}{*}{3C 105} & \multirow{2}{*}{2006-11-21} & \multirow{2}{*}{0.48} & \multirow{2}{*}{0.92}& \multirow{2}{*}{41.15} & \multirow{2}{*}{40.26}& \multirow{2}{*}{41.5} & \multirow{2}{*}{17.38} & \multirow{2}{*}{44.21}  & \multirow{2}{*}{513} \\
 & & & & & & & & & \\
\multirow{2}{*}{J0318.7+6828} & \multirow{2}{*}{2006-11-21} & \multirow{2}{*}{0.72} & \multirow{2}{*}{0} & \multirow{2}{*}{41.17} & \multirow{2}{*}{40.74} & \multirow{2}{*}{41.64} & \multirow{2}{*}{7.94} & \multirow{2}{*}{44.08}  & \multirow{2}{*}{275} \\
 & & & & & & & & & \\
\multirow{2}{*}{J0918.5+0425} & \multirow{2}{*}{2004-03-09} & \multirow{2}{*}{0.04} & \multirow{2}{*}{0.26} & \multirow{2}{*}{41.63} & \multirow{2}{*}{41.03} & \multirow{2}{*}{42.1} & \multirow{2}{*}{14.75} & \multirow{2}{*}{44.30}  & \multirow{2}{*}{159} \\
 & & & & & & & & & \\
\hline
\end{tabular}
\end{minipage}
\end{table*}%
All sources show a secondary power law arising in the soft X-ray band. A prominent soft X-ray emission is a common feature of optically selected samples of Seyfert 2 galaxies, with a scattering fraction in the range 3\%-10\% (\citealt{tur97}, \citealt{cap06}, \citealt{gua05}). Follow-up Suzaku observations of Swift/BAT AGN led to the discovery of a new type of AGN (``buried'' or ``hidden'' AGN) with an extremely low scattering fraction (0.5\%, see \cite{ued07}, \cite{egu09}). A quarter of 32 AGN analyzed by \cite{nog10} from the XMM-Newton Serendipitous Source catalog presents scattering fractions below 0.5\%. W09a found that hidden AGN constitute a high percentage of the BAT AGN sample ($\sim24\%$), adopting the criteria that a hidden source is one where the scattering fraction (f$_{\rm scatt}$) is $\leq3\%$ and the ratio of soft (0.5--2 keV) to hard (2--10 keV) observed flux is $\leq4\%$. This subsample of hidden AGN also includes 3C 105 and J0918.5+0425, while 3C 452 and J0318.7+6828 are not part of this class because they have a f$_{\rm scatt}>3\%$. 
\\
We confirm the W09a findings for the first two sources, but on the basis of the Suzaku and XMM-Newton analysis, we infer an upper limit of $0.5\%$ to the scattering fraction for 3C 452, probably because W09 did not model the reflection component. The scattering fraction of J0318.7+6828 is variable around the classification limit. Considering that the W09 hidden AGN show an averaged column density N$_{\rm H}>3\times10^{23}$ cm$^{-2}$, about an order of magnitude higher than the value shown
by J0318.7+6828, the object with the lowest column density in our sample, we do not classify J0318.7+6828 as a ``hidden'' AGN.
\\
It is thought that the soft X-ray emission is most often due to scattered light by the photo-ionized gas in the opening part of the torus while the direct view to the nucleus is absorbed, and the relative intensity of the scattered component with respect to the intrinsic one can be related to the opening angle of the optically thick torus. This new class of objects (\citealt{ued07}) is deeply buried with a small opening angle torus ($<20^{\circ}$ for the most extreme cases) and viewed in a face-on geometry.
As a consequence, the same object seen edge-on would appear as a heavily Compton-thick AGN (N$_{\rm H}\sim10^{25}$ cm$^{-2}$). Only 3C 452 of our sample, which is strongly absorbed, reflection dominated and has a $<0.5\%$ scattering fraction, could effectively fall within this new class of AGN.
However, the three radio galaxies of our sample could be contaminated in the soft X-ray emission by the base of the jet in the unresolved nuclear region, so that our findings would only pose an upper limit to the scattering fraction (\citealt{2009MNRAS.396.1929H}).
\\
Since the photo-ionized gas responsible for the scattered light is probably the same region that generates the [OIII] emission optical lines (\citealt{bia06}) of the Narrow Line Region (NLR), a low torus opening angle could affect the optical band. This bias would explain the lack of hidden AGN in optically selected samples. \cite{2010ApJ...710..503W} (W10 hereafter) listed the optical properties of the BAT AGN sample, comparing the absorption-corrected [OIII]$\lambda5007$ with the BAT 14--195 keV luminosity. Contrary to the results of \cite{hec05}, they found only a weak linear correlation between the reddening-corrected [OIII] luminosity and L$_{\rm X}$. They suggested that the [OIII] luminosity is affected by extinction poorly corrected by using the Balmer decrement, and thus it cannot be used as an unbiased and accurate tracer of the AGN power. Similar results were obtained by \cite{nog10} who found R$^{\rm X}_{\rm [OIII]}>10$ (the ratio of the intrinsic 2--10 keV power-law luminosity with respect to the corrected L[OIII]) in their sample of $<3\%$ scattering fraction AGN and a value $>100$ for the most extreme cases, which is higher than the 1--100 range that they found from a subsample of Seyfert 2 from \cite{bas99} selected in the same range of N$_{\rm H}$ ($0.6-20\times10^{23}$ cm$^{-2}$). A strong test for the buried nature of our targets would be a high value of R$^{\rm X}_{\rm [OIII]}$.
\\
In Table \ref{tab:opt} the extinction-corrected luminosity of the main optical emission lines of our sample is listed as given by W10, along with the Galactic and the intrinsic reddening applied, the latter as inferred from the Balmer decrement. The last column shows the ratio between the intrinsic 2--10 keV power law luminosity and the absorption-corrected [OIII] luminosity (R$^{\rm X}_{\rm [OIII]}$). All ratios are well above the 1--100 range of the \cite{bas99} sample, and agree very well with the values of L$_{\rm [OIII]}$/L$_{\rm X}$ for narrow line objects reported in W10. The scatter in R$^{\rm X}_{\rm [OIII]}$ in the sample is only a factor of 5, with 3C 452, the AGN with the lowest scattering fraction and a coherent hard X-ray luminosity, reaching the maximum with a value of R$^{\rm X}_{\rm [OIII]}\sim800$, which is close to the extreme cases of \cite{nog10}. 
\\
The clearest result from the sample optical properties is that the X-ray to [OIII] luminosity ratio follows the X-ray properties of our sources. The three radio galaxies present an increasing R$^{\rm X}_{\rm [OIII]}$ as the column density increases and the secondary, scattered emission at soft X-rays becomes dimmer.
Although the sample is far from being complete, the [OIII] luminosity is still affected by extinction, according to the hypothesis of W10, and it is a biased tracer of the AGN power, with a bias level depending on the geometry of the surrounding matter.
The lowest luminosity ratio indeed comes from the most peculiar object: the QSO-like, highly variable J0918.5+0425, the only source that does not emit in the radio band and shows both the highest X-ray and [OIII] luminosity. 
\\
Our findings seem to confirm the origin of the low 3C 452 [OIII] relative luminosity as induced by a clumpy, small opening angle absorber, blocking the emission of the NLR where the [OIII] lines arise, which proves the hidden nature of 3C 452.

\section{Radio connection}
   \begin{figure}[!h]
   \centering
   \includegraphics[width=.45\textwidth, angle=0]{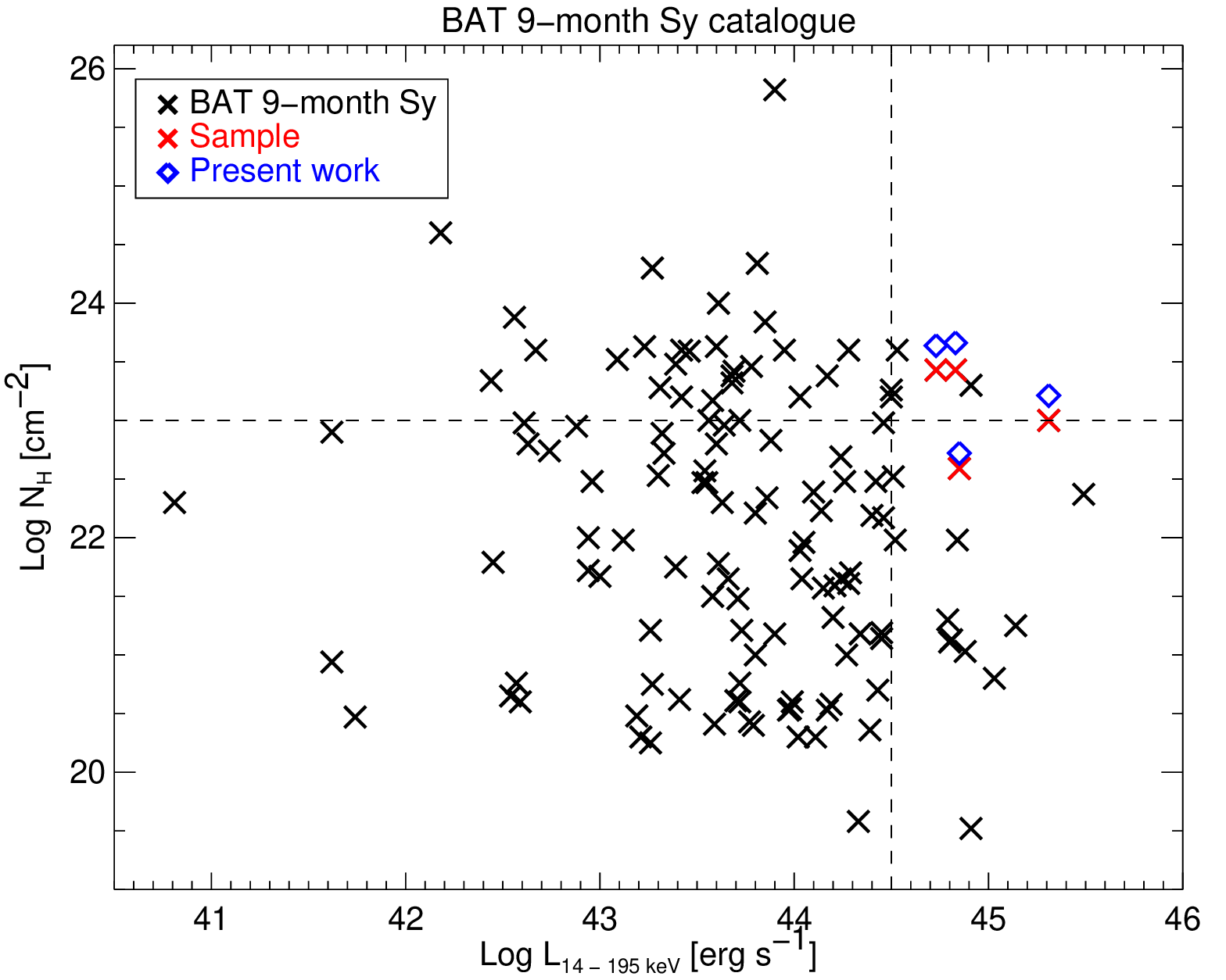}\\
   \includegraphics[width=.45\textwidth, angle=0]{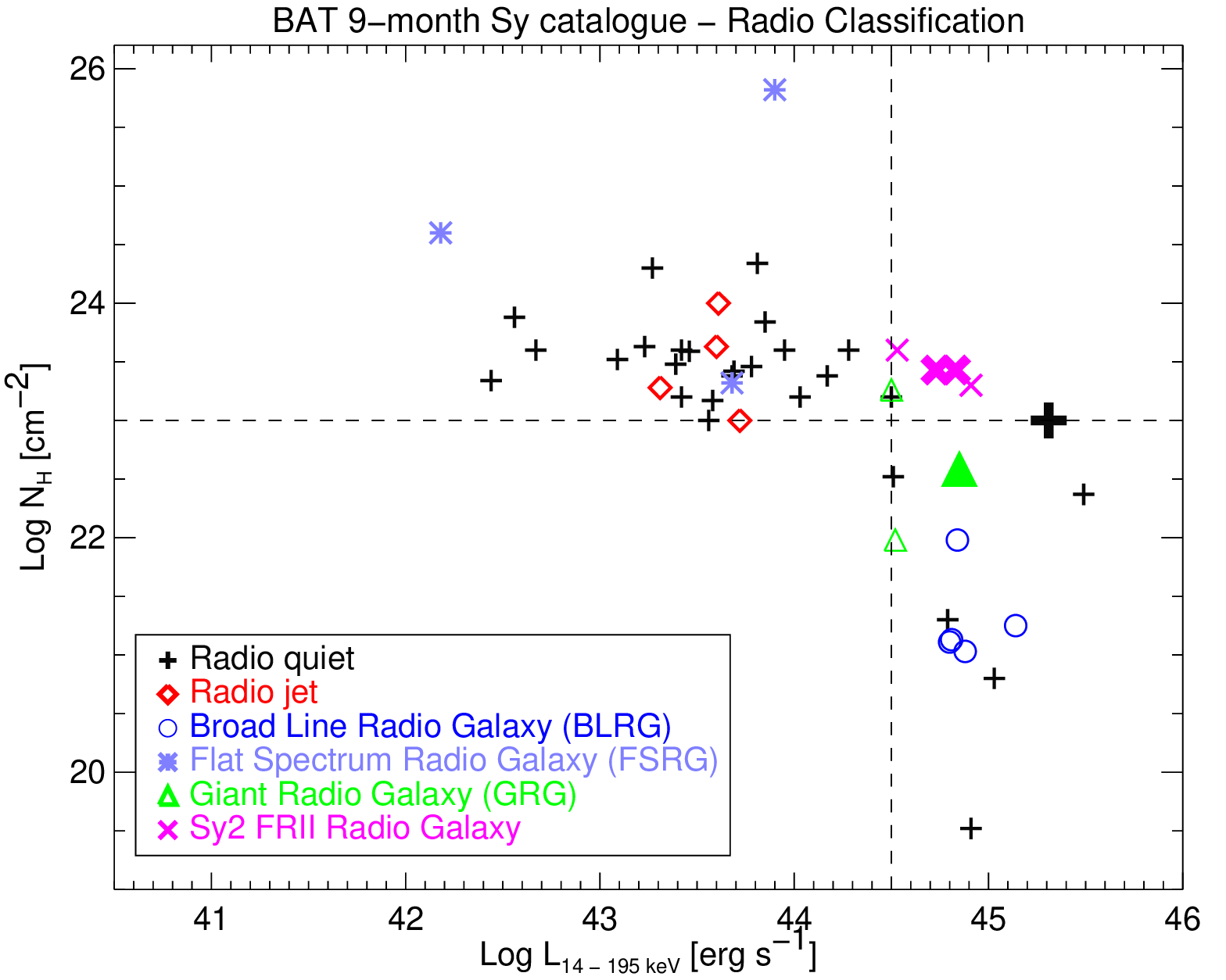}
   \caption{\textit{Top panel}: The distribution in column density (N$_{\rm H}$) and BAT 14--195 keV luminosity of the BAT nine-month survey Seyfert galaxies as given by \cite{tue08}. The red crosses and blue diamonds represent our sample column density from the BAT survey and the Suzaku analysis. \textit{Bottom panel}: The BAT nine-month Seyferts subsample is defined by Log N$_{\rm H}\geq23$ and/or Log L$_{14--195 keV}\geq44.5$ (dashed lines), i.e., the most luminous and absorbed sources are selected. The symbols refer to the radio type as classified from the literature, the thicker symbols indicate the present sample.\label{fig:cross}}%
    \end{figure}
As mentioned in the introduction, the amount of obscuration in AGN is inversely proportional to the X-ray luminosity (\citealt{gil07}). Despite this observational evidence, our sources are highly luminous and highly absorbed. Here we examine the possibility that their high column density could be related to their radio properties.
\\
In the sample, the most absorbed sources (3C 452 and 3C 105) are powerful FRII radio galaxies, J0318.7+6828 is also a giant radio galaxy. The only source that is an exception from our assumption is J0918.5+0425, which is a highly variable QSO removed from the BAT 9-month survey, however.
\\
We took into account the BAT nine-month catalog, and, after selecting the Seyfert galaxies, i.e., discarding the Blazar/BL-LAC/Galaxy sources, the column density as a function of the BAT 14--195 keV luminosity is plotted in Fig. \ref{fig:cross} (top panel), using the values reported by \cite{tue08}. The crosses refer to the BAT Seyfert catalog, the red crosses represent the present sample, while the blue diamonds show the column density obtained from the Suzaku analysis. We assumed, as reported by W09a, a proportional relation between the BAT 14--195 keV and the softer Suzaku 2--10 keV luminosity.
\\
Evidently, our sources are at the edge of the N$_{\rm H}$ - L$_{\rm X}$ relation, i.e., they are both the most absorbed and the most luminous of the BAT Seyfert catalog. If this peculiar feature is related to the radio power, we should not find sources similar in terms of radio properties in other regions of the plot. To investigate this possibility, we traced two arbitrary regions (see dashed lines in Fig. \ref{fig:cross}) that define the most absorbed (Log N$_{\rm H}\geq23$) and/or the most luminous (Log L$_{14-195 \;\rm keV}\geq44.5$) sources of the Seyfert sample. At the intersection we find our sources, except for J0318.7+6828 (the lowest diamond) which is, in fact, a Compton-thin AGN with a column density of N$_{\rm H}\sim5\times10^{22}$ cm$^{-2}$. We searched the literature for the radio classification of the BAT sources selected in terms of high luminosity/column density, which was difficult because many sources are defined differently according to different authors. The BAT Seyferts are divided into: i) radio quiet (including sources with a low radio flux that does not permit a radio classification); ii) radio jet (sources with a jet, but without a radio type/morphology definition); iii) broad line radio galaxy (BLRG); iv) flat spectrum radio galaxy (FSRG); v) giant radio galaxy (GRG); vi) Seyfert 2 FRII radio galaxy. The result is plotted in Fig. \ref{fig:cross} (bottom panel) and the thicker symbols refer to our sample.
\\
It is not surprising to find out that different classes of radio galaxies are grouped differently according to the X-ray luminosity and the amount of absorbing matter (see e.g. \citealt{1999ApJ...526...60S}), because the radio morphology depends on the inclination of the AGN with respect to the observer and is connected to the radio power. 
For example, the column density increases from BLRG (blue circles) to GRG (green triangles) and, finally, FRII Radio Galaxies (pink crosses). However, we note that, in addition to 3C 452 and 3C 105, and considering the GRG as a different class of objects, there are only two other similar FRII galaxies in the selection, and both lie in the high-luminosity/absorption region. The only three GRG, including J0318.7+6828, share the luminosity range with the classic FRII, but they are all Compton-thin AGN. Radio-quiet sources, such as QSO J0918.5+0425, are instead uniformly distributed in the N$_{\rm H}$ - L$_{\rm X}$ plot.
\\
Given these findings, it is clear that our sources are four of the most luminous and obscured AGN of the BAT survey because three of them are powerful FRII radio galaxies, while J0918.5+0425 was confirmed in its peculiar nature. The Suzaku/XIS angular resolution does not allow us to separate the potential X-ray jet from the AGN core itself. However, if an X-ray emission enhancement were caused by the jet, the effect on the scattering fraction would be an overprediction of the parameter, which in turn would consolidate the hidden nature of 3C 452. A higher X-ray luminosity could also cause an increase of the X-ray/[OIII] luminosity ratio R$^{\rm X}_{\rm [OIII]}$. While such an effect cannot be excluded, it must be noted that the highest R$^{\rm X}_{\rm [OIII]}$ is associated to the lowest 2--10 keV intrinsic luminosity (3C 452).

\section{Summary and discussion}\label{sec:disc}
We studied four of the most luminous BAT AGN ($44.73 < \rm{Log\;L}_{\rm BAT} < 45.31$): J2246.0+3941 (3C 452), J0407.4+0339 (3C 105), J0318.7+6828, and J0918.5+0425.
Of these, 3C 452 is the only certain Compton-thick AGN candidate because of four properties:
\begin{enumerate}
\item The source spectrum shows both high absorption (N$_{\rm H}\sim4\times10^{23}$ cm$^{-2}$) and strong Compton reflection, which also flattens the intrinsic power-law photon index. The relative strength of the reflection is consistent with a complex absorber where a large portion of the subtended solid angle is composed by gas that is optically thicker than the amount crossed along the line of sight;
\item The XIS and PIN light curves show no significant short-term variability, and the spectral parameters of the source stay the same during six years. The lack of variability is a trace of reflection-dominated emission, because the nuclear intrinsic variability is diluted by radiation coming from the torus, which covers a much more extended region.
\item Although the soft X-ray band is contaminated by the extended thermal gas and the lobes' IC emission, a low scattering fraction ($<0.5\%$) was found, confirmed by the XMM observation for which the lower PSF size allowed us to extract only the AGN region. This aspect, coupled with an extremely high ratio of the intrinsic 2--10 keV to the [OIII] luminosity, classifies 3C 452 as hidden AGN, where, as reported in Sec. \ref{sec:opt}, the source is seen face-on and the equatorial region is covered by Compton-thick matter. This scenario would be consistent with the hypothesis of a complex absorber.
\item The EW of the fluorescence iron line is lower than expected for a Compton-thick AGN and consistent with the evaluated column density, for an inclination angle $>60^{\circ}$ (\citealt{ghi94}, \citealt{ike09}). However, such a low EW could be justified by a Fe abundance lower than the solar value, in particular, an EW = 164 eV is obtained for a Fe abundance in the range 0.2-0.3 (\citealt{bal02}). The quality of the data only allowed us to evaluate an upper limit to the Fe abundance of 0.7. It should be noted that supersolar metal abundances are expected for broad line Seyfert 1s (\citealt{2001KOMOSSA}).

\end{enumerate}
These properties assume more importance when comparing 3C 452 with 3C 105. Despite their comparable column density, redshift, X-ray luminosity, and radio properties, 3C 105 shows a long-term variability and is not reflection-dominated, while 3C 452 is. These differences underline the necessity of hard X-ray observations to better characterize the absorbing medium and to find sources.
\\
In contrast, the overall properties of 3C 105 lead us to classify this source as a Compton-thin AGN.
The most prominent feature is the strong variability in flux and scattering fraction found by comparing our results with the Chandra, XMM-Newton, and XRT observations. The flux variability is also confirmed by the high normalization factor between the PIN and BAT data ($\sim1.7$) and the slight variations in the Suzaku light curves. A luminosity variability, especially on time scales of months/years, is not surprising for Type II AGN not dominated by scattering, because the moderate column density in the line of sight only blocks the soft X-ray energy range ($\rm E<3$ keV). A change in the scattered/partially covered emission is instead very peculiar, and, coupled with the flux changes, it could reveal the distance and extension of the scattering region. Among the four observations, the highest scattering component ($\sim1.4\%$) was found in the XIS detection, which also needs a more extended extracting region. This could imply that we are including in the source model some external component. However, no external point sources or extended thermal emission are detected from the analysis of the high-resolution Chandra/ACIS image, while a possible contamination of the southern hotspot would be below the scattering factor value, considering also the statistical uncertainties (see Sec. \ref{3c105_fit_res}). In addition, the analysis of the XIS spectrum of 3C 105 reported in W08 returns a higher scattering fraction ($\sim2.7\%$). If the scattering component is intrinsically variable, it can be interpreted differently according to the proposed models: 
\begin{enumerate}
\item The arising power law in the soft band is nuclear emission escaping through a partially covered medium. A change in its intensity can be associated to the motion of the clouds along the line of sight, according to the way \cite{2002ApJ...571..234R} modeled the variability in the absorbing column density. Extending the computation of these authors to this case and using the 3C 105 black hole mass as given by W10, a $\Delta\rm N_{\rm H}=5\times10^{23}$ and a time delay of 50 days (Chandra/Suzaku observations time lapse) leads to an absorber distance $<1$ pc for a density $\rho<10^{9}$ cm$^{-3}$.
\item The soft emission is scattering from a distant region (e.g., the NLR). A change in the nuclear flux results in a change of the scattered emission, with a time delay given by the distance of the scattering region from the nucleus. An increase in luminosity can also increment the ionization of the surrounding medium, increasing the scattering efficiency. 
\end{enumerate}
Because we did not find absorption in the scattering component, it is unlikely that a fraction of the highly obscured intrinsic emission can arise from the inner regions of the torus without absorption. On the other hand, the second hypothesis correlates well with the high variability of the absorption-corrected X-ray luminosity and the low L$_{\rm 2-10 \;\rm keV}$/[OIII] ratio. 
\\
The same variability behavior is found in J0318.7+6828, although weaker. This AGN is the least absorbed of our sample (N$_{\rm H}\sim4\times10^{22}$ cm$^{-2}$), and the only one not classified as a hidden AGN. The scattering fraction is variable by around 3\%, a change that correlates with a 30\% variability in the 2--10 keV band. Compared with 3C 105, the scattering fraction from the XIS analysis is about half of the value from the XMM-Newton observation, although the first requires a wider extracting region. 
We conclude that the variability of the soft scattered emission seems a ubiquitous property of these two sources, but a dedicated X-ray monitoring for a longer time is needed to effectively relate these findings to the geometry and composition of the scattering/partially covering region.
\\
Finally, the farthest source of our sample, J0918.5+0425, is a moderately absorbed Type II AGN. The strong variability that led this source to be removed from the 22-month BAT survey is confirmed from comparing the XRT, Suzaku and nine-month-averaged BAT data. However, longer exposures and a finer spatial resolution are needed for a more detailed analysis.
\\
Last but not least, some considerations on the absolute luminosity of the sample. The definition of a Type II QSO is still arbitrary: according to \cite{zak03}, the AGN requires a Log L[OIII]$\lambda5007>40.06$ erg s$^{-1}$ to be classified as QSO, while the QSO sample of \cite{kru08} is based on absorption-corrected, intrinsic 0.5--10 keV luminosities $>10^{44}$ erg s$^{-1}$.
The only source to satisfy both definitions is J0918.5+0425, which is confirmed to be a Type II QSO as reported in \cite{tue08}. However, as pointed out in Sec. \ref{sec:opt}, the [OIII] luminosity could suffer from extinction in case of high absorbing column densities. If we only consider the X-ray classification, then 3C 105 and J0318.7+6828 can also be defined as QSO sources.
\\
The common properties of all sources, except J0918.5+0425, are that they show QSO-like luminosity, are obscured, present low scattering fractions and are powerful FRII radio galaxies, the latter playing a key role in the luminosity-absorption relation. Is this a secondary effect due to the unresolved jet base contaminating the X-ray luminosity? Or is there an intrinsic, mutual connection between the jet production mechanism and the matter surrounding the SMBH? Given the limited number of sources, it is impossible to reveal the nature of this connection here, or if the decreasing number of Compton-thick objects at higher luminosities is related to the radio population at these energies. In this context, the increase in the AGN survey accuracy in terms of completeness and energy coverage, as proven by the Swift/BAT survey and follow-up optical and soft X-ray catalog, along with the improvement in the sensitivity and spectral resolution of hard X-ray observations (e.g. NuStar), will be a fundamental tool to unfold the AGN cosmic history. 


\begin{acknowledgements}
We thank the anonymous referee for useful suggestions to improve
the paper. This work was supported by the Thales Alenia Space Italy Ph.D. fellowship and the University Space Research Association (USRA), and it was hosted by the Center for Research and Exploration in Space Science and Technology (CRESST, Contract NNG06EO90A) and the NASA/GSFC. V.F. gratefully acknowledges G.G.C. Palumbo for his constant scientific support and P. Grandi for her extensive paper reading.

\end{acknowledgements}

\bibliographystyle{aa}
\bibliography{fioretti_suzaku}

\end{document}